\documentclass[fleqn,usenatbib]{mnras}

\pdfoutput=1
\usepackage{newtxtext,newtxmath}

\usepackage[T1]{fontenc}

\DeclareRobustCommand{\VAN}[3]{#2}
\let\VANthebibliography\thebibliography
\def\thebibliography{\DeclareRobustCommand{\VAN}[3]{##3}\VANthebibliography}

\usepackage{array}
\usepackage{hanging}
\usepackage{caption}
\usepackage{subcaption}
\usepackage{lipsum}  
\usepackage{xcolor}

\usepackage{graphicx}	
\usepackage{amsmath}	
\usepackage{booktabs}

\title[Environmental quenching at $z\sim1$]{Insights into environmental quenching at $z\sim1$: an enhancement of faint, low-mass passive galaxies in clusters}

\author[H. Gully et al.]{Harry Gully$^{1}$\thanks{E-mail: ppyhg3@nottingham.ac.uk},
Nina Hatch$^{1}$, Syeda Lammim Ahad$^{2,3}$, Yannick Bah\'e$^{1,4}$, Michael Balogh$^{2,3}$, Devontae C. Baxter$^{5}$, \newauthor Pierluigi Cerulo$^{6}$, M. C. Cooper$^{7}$, Ricardo Demarco$^{8}$, Ben Forrest$^{9}$, Umberto Rescigno$^{10}$, Gregory Rudnick$^{11}$, \newauthor Benedetta Vulcani$^{12}$, Gillian Wilson$^{13}$
\\
$^{1}$School of Physics and Astronomy, University of Nottingham, Nottingham NG7 2RD, UK\\
$^{2}$Department of Physics and Astronomy, University of Waterloo, Waterloo, ON N2L 3G1, Canada\\
$^{3}$Waterloo Centre for Astrophysics, University of Waterloo, Waterloo, ON N2L 3G1, Canada\\
$^{4}$Laboratory of Astrophysics, \'{E}cole Polytechnique F\'{e}d\'{e}rale de Lausanne (EPFL), Observatoire de Sauverny, 1290 Versoix, Switzerland\\
$^{5}$Department of Astronomy \& Astrophysics, University of California, San Diego, 9500 Gilman Dr, La Jolla, CA 92093, USA \\
$^{6}$Departamento de Ingenier\'ia Inform\'atica y Ciencias de la Computaci\'on, Facultad de Ingenier\'ia, Universidad de Concepci\'on, Casilla 160-C, Concepci\'on, Chile\\
$^{7}$Department of Physics \& Astronomy, University of California, Irvine, 4129 Reines Hall, Irvine, CA 92697, USA\\
$^{8}$Institute of Astrophysics, Facultad de Ciencias Exactas, Universidad Andr\'es Bello, Sede Concepci\'on, Talcahuano, Chile \\
$^{9}$Department of Physics and Astronomy, University of California, Davis, One Shields Avenue, Davis, CA 95616, USA\\
$^{10}$Instituto de Astronom\'ia y Ciencias Planetarias de Atacama (INCT), Universidad de Atacama, Copayapu 485, Copiap\'o, Chile\\
$^{11}$Department of Physics \& Astronomy, The University of Kansas, Malott Room 1082, 1251 Wescoe Hall Drive, Lawrence, KS 66045, USA\\
$^{12}$INAF-Osservatorio astronomico di Padova, Vicolo Osservatorio 5, I-35122 Padova, Italy\\
$^{13}$Department of Physics, University of California Merced, 5200 North Lake Road, Merced, CA 95343, USA
}

\date{Accepted XXX. Received YYY; in original form ZZZ}

\pubyear{2025}

\begin{document}
\label{firstpage}
\pagerange{\pageref{firstpage}--\pageref{lastpage}}
\maketitle

\begin{abstract} 
Understanding the processes that transform star-forming galaxies into quiescent ones is key to unraveling the role of environment in galaxy evolution. We present measurements of the luminosity functions (LFs) and stellar mass functions (SMFs) of passive red-sequence galaxies in four galaxy clusters at $0.8 < z < 1.3$, selected using deep VLT observations complemented with data from the GCLASS and GOGREEN surveys. We find a significant enhancement in the abundance of faint/low-mass passive galaxies in both the LFs and SMFs of all four clusters compared to the field. This is further evidenced by a shallower low-mass slope in the composite passive cluster SMF, which yields a Schechter parameter $\alpha = -0.54^{+\,0.03}_{-0.03}$, compared to $\alpha = 0.12^{+\,0.01}_{-0.01}$ for the field. Our findings indicate that quenching processes that act in clusters are enhanced compared to the field, suggesting that environmental quenching mechanisms may already be active by $z\sim1$. To reproduce the observed passive cluster SMF, we estimate that $25\pm5\%$ of the star-forming field population that falls into the cluster must have been quenched. Our results largely support traditional quenching models but highlight the need for deeper studies of larger cluster samples to better understand the role of environmental quenching in the distant Universe.
\end{abstract}




\section{Introduction}

The cessation of star formation in galaxies leads to a distinct bimodality in the galaxy population \citep[e.g.][]{Strateva2001, Kauffmann2003, Baldry2004, Bell2004, Brammer2011}. Those still forming new stars are bluer in colour and will typically have disk-like morphologies. Conversely, quiescent galaxies are redder with more spheroidal morphologies, and will have little-to-no star formation. A number of processes have been proposed to explain the cessation of star formation, falling into two main categories: mass quenching and environmental quenching. The former term, introduced by \cite{Peng2010}, describes quenching processes that correlate with the stellar mass of a galaxy while the latter term describes processes that correlate with environment.

It is thought that mass quenching is driven by feedback from active galactic nuclei \citep{Bower2006, Croton2006, Bremer2018}, or gas outflows from stellar winds or supernovae explosions \citep{Larson1974, Dekel1986, Dalla2008, Oppenheimer2010}. Environmental quenching, however, is thought to occur as a galaxy accretes onto a large halo, upon which it can be stripped of its gas through tidal or ram-pressure stripping \citep{Gunn1972}, be prevented from accreting hot gas, known as strangulation \citep{Larson1980}, or undergo harassment or mergers \citep{Moore1996}. These processes are more effective in dense cluster environments due to the high relative velocities\footnote{Except for mergers, which are more common in groups as the high relative velocities in clusters make mergers less likely.} and dense intracluster medium, which can more readily strip or heat a galaxy's gas. Altogether, these environmental mechanisms can suppress the formation of new stars, transforming once highly star-forming galaxies into red, dead, passive galaxies. Some studies of the low-redshift Universe ($z<1$) suggest that mass and environmental quenching processes are generally independent \citep[e.g.][]{Baldry2006, Peng2010, Peng2012, Kovac2014, vanderBurg2018}. Others, however, provide alternative interpretations in which the hierarchical formation of structure necessitates that the two processes are intertwined and cannot be separated \citep[e.g.][]{delucia2012, Contini2020}. At even earlier epochs ($z>1$), there is evidence that independence between the two processes does not hold \citep{Balogh2016, Darvish2016, Kawinwanichakij2017, Papovich2018, Pintos2019, vanderBurg2020, McNab2021}. It is also unclear whether dynamical processes such as tidal or ram-pressure stripping dominate at higher redshifts like they do in the local Universe \citep{Boselli2014, Boselli2022, Fossati2016, Bellhouse2017, Zinger2018, Jaffe2018, Foltz2018, Cramer2019, Moretti2022, Vulcani2022}. 

One of the more puzzling results of the last decade is the lack of distinction between the shapes of the SMFs of cluster and field passive galaxies at $z\sim1$ \citep{vanderBurg2013, vanderBurg2020}. According to traditional models at lower redshifts, environmental quenching is expected to operate independently of stellar mass. Due to the abundance of low-mass star-forming galaxies that accrete onto clusters, a signature of environmental quenching is a relative upturn in the SMF of passive cluster galaxies towards the low-mass end \citep[e.g.][]{Weinmann2006, Boselli2006, Baldry2006, Muzzin2013}. In the absence of environmental quenching (i.e. in the field), the SMF of passive galaxies does not exhibit such an upturn. In contrast, \cite{vanderBurg2013} and \cite{vanderBurg2020} find almost identical SMF shapes between passive cluster and field galaxies at $z\sim1$. This would seem to suggest a complete lack of environmental quenching if it were not for the overall enhancement in the passive galaxy fraction in clusters. \cite{Tomczak2017} also find no such upturn at low masses, but do find a higher ratio of high-mass to low-mass galaxies in clusters, similar to results from protoclusters \citep{Forrest2024}. These discrepancies imply that environmental quenching may not operate in the same way it does at lower redshifts, and instead has some stellar mass dependence. Whether this is caused by a weaker version of environmental quenching at higher redshifts, or a fundamentally different process altogether remains unclear. 

This challenge to traditional environmental quenching models can be alleviated if we assume that the relative upturn in the SMF of passive cluster galaxies compared to the field occurs at even lower masses, where environmental quenching would primarily affect galaxies below the mass limits of previous studies. In this work, we extend the GCLASS-based (see section~\ref{ClusterSample} for details) study of \citet[which reaches down to $10^{10}$ M$_{\odot}$]{vanderBurg2013} to fainter magnitudes and lower masses using deep VLT observations of four $0.8 < z < 1.3$ galaxy clusters, where we measure the LF and SMF of passive galaxies. Probing the lower masses where model predictions are most discrepant \citep[e.g.][]{Guo2011, Weinmann2012, Bahe2017, Lotz2019, Kukstas2023} will help us to determine whether fundamentally different physics is needed to explain observations.

The structure of the paper is as follows: In section~\ref{Data}, we give an overview of the photometric data used to compile our cluster sample, as well as a description of the field sample used as a reference. Section~\ref{Analysis} presents our red-sequence selection, measurements of rest-frame colours and stellar masses, and the method for constructing the LFs/SMFs. In section~\ref{Results}, we present our results, specifically comparing the LFs/SMFs between the cluster and field environments. We discuss these results in section~\ref{Disc} in the context of environmental quenching processes, with our conclusions given in section~\ref{Conc}.

Unless stated otherwise, the halo mass definition we adopt is the mass enclosed by a sphere that has a density 200 times the critical density of the Universe (M$_{200}$), with its corresponding R$_{200}$ radius. All magnitudes are given in the AB system. We adopt $\Lambda$CDM cosmology with $\Omega_0=0.3$, $\Omega_{\Lambda}\,=\,0.7$, and $H_0 = 70$ km s$^{-1}$ Mpc$^{-1}$.

\section{Data}
\label{Data}

\begin{figure}   \includegraphics[width=\columnwidth]{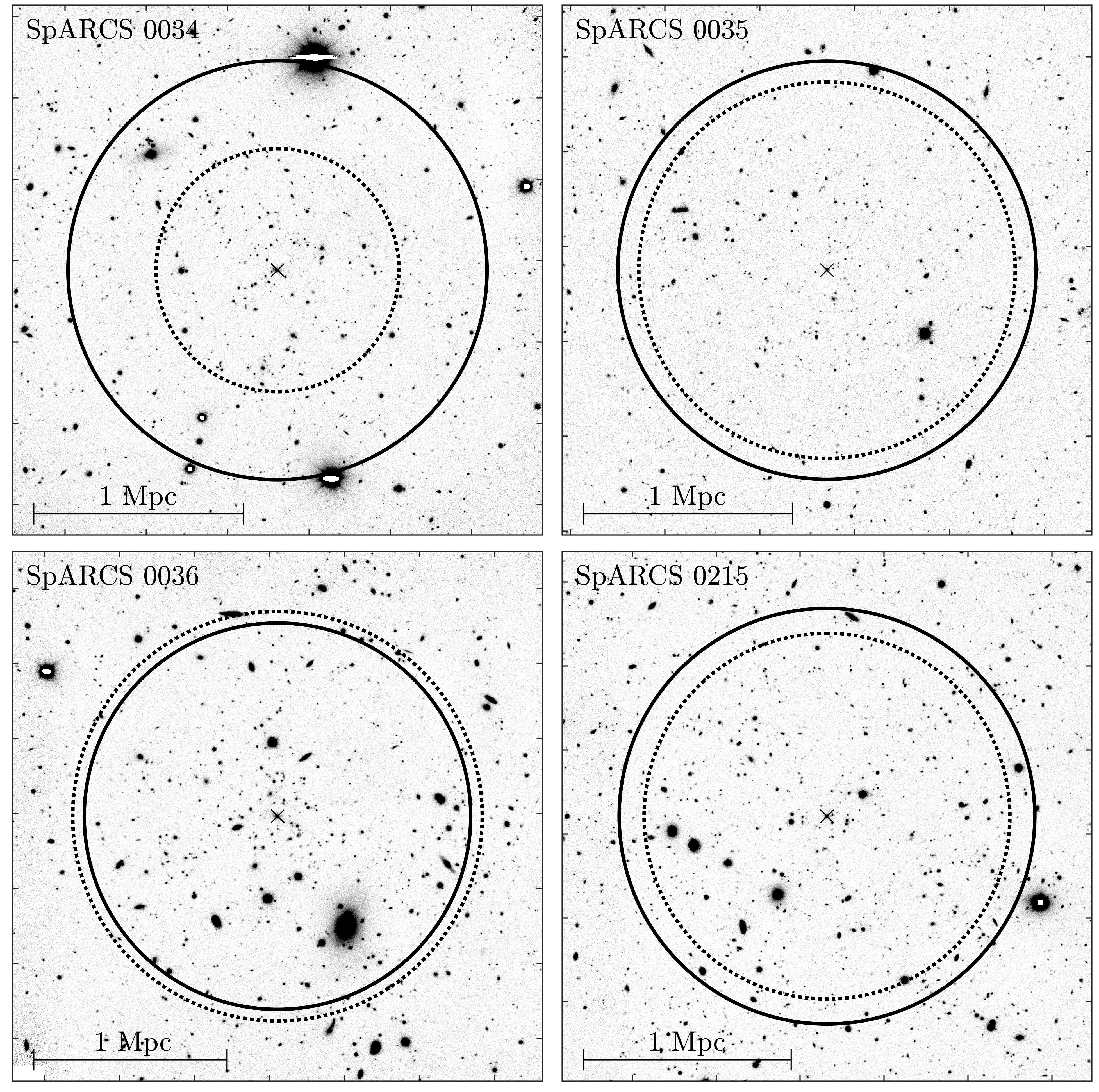}
    \caption{\small{The VLT detection-band images ($F_r$) of the four clusters. The black cross marks the centre of the cluster, while the black dotted circle represents R$_{200}$ (values given in Table~\ref{Clusterinfo}). The solid black circle has a radius of 1 Mpc, and is used to separate the inner and outer regions that are used in the statistical background subtraction in section~\ref{SBS}. These images are focused on the cluster centres and do not represent the full observed regions.}}
        \label{TCIms}
\end{figure}

\begin{table*}
\centering
\caption{GCLASS cluster properties.}
\begin{tabular}{wl{1.7cm} wc{0.9cm} wc{1cm} wc{0.9cm} wc{1.2cm} wc{1.2cm} wc{1cm} wc{0.65cm} wc{0.65cm} wc{1cm} wc{1.2cm} wc{1.3cm}}
\hline
Name & RA & Dec & $z$ & $\sigma_v$$^\textrm{\textit{a}}$ & M$_{200}$$^\textrm{\textit{a}}$ & R$_{200}$$^\textrm{\textit{a}}$ & \multicolumn{2}{c}{Filters$^\textrm{\textit{b}}$} & [$F_r$]$_{\textrm{lim}}$$^\textrm{\textit{c}}$ & \multicolumn{2}{c}{Completeness Limits$^\textrm{\textit{d}}$} \\
& \multicolumn{2}{c}{(J2000)} & & (km s$^{-1}$) & ($10^{14}$ M$_{\odot}$) & (Mpc) & $F_b$ & $F_r$ & ($5\sigma$, AB) & Mag (AB) & Mass (M$_{\odot}$) \\
\hline
SpARCS 0034 & 8.6751 & $-43.1315$ & 0.867 & $405\pm51$ & $0.6\pm0.2$ & 0.58 & 691\,nm & 834\,nm & 24.74 & 24.91 & $2.27\times10^9$\\
SpARCS 0035$^\textrm{\textit{e}}$ & 8.9571 & $-43.2068$ & 1.335 & $840\pm111$ & $3.8\pm1.5$ & 0.90 & 834\,nm & $Y$ & 25.16 & 24.41 & $9.07\times10^9$\\
SpARCS 0036 & 9.1875 & $-44.1805$ & 0.869 & $799\pm82$ & $3.6\pm1.1$ & 1.06 & 691\,nm & 834\,nm & 24.37 & 24.46 & $3.18\times10^9$\\
SpARCS 0215 & 33.8500 & $-3.7256$ & 1.004 & $656\pm70$ & $2.4\pm0.8$ & 0.88 & 691\,nm & 815\,nm & 24.73 & 24.83 & $6.59\times10^{9}$\\
\hline
\end{tabular}
 \small
 \begin{flushleft} \textit{\textbf{Notes}}. $^\textrm{\textit{a}}$The velocity dispersions, halo masses and radial scales are from \cite{Biviano2021}, which were obtained using the MAMPOSSt method \citep{Mamon2013}. We note here that the halo masses do not seem to correlate with cluster richness. However, the halo masses are derived from the velocity dispersions which have fairly large uncertainties. It is therefore likely that the masses of each cluster are similar.\\
 $^\textrm{\textit{b}}$These filter pairs are chosen to span the 4000\,{\AA} break of galaxies at the respective cluster redshift, where $F_b$ corresponds to the bluer filter and $F_r$ the redder one. The 691\,nm, 815\,nm and 834\,nm filters are the `special' intermediate-band FORS2 filters (aka the night sky suppression filters), while $Y$ is the broadband HAWK-I filter centered at 1020\,nm. See section~\ref{VLTObs} for more details of the observations. \\
 $^\textrm{\textit{c}}$The $5\sigma$ limits quoted are measured using $2\arcsec$ apertures. See section~\ref{VLTObs} for details of this measurement.\\
 $^\textrm{\textit{d}}$The $70\%$ completeness limits of galaxies, given in both magnitude and mass. The limits in magnitude come directly from the injection-recovery simulation (see section~\ref{compcorr}), while those in mass are converted from magnitude using the mass-luminosity relation derived in section~\ref{RFcol}, and only correspond to galaxies at the redshift of the cluster. \\
 $^\textrm{\textit{e}}$Cluster also in the GOGREEN survey.\\ 
 \end{flushleft}
 \label{Clusterinfo}
\end{table*}

\subsection{Cluster sample}
\label{ClusterSample}

The four clusters studied in this paper are a subsample of clusters observed in the Gemini CLuster Astrophysics Spectroscopic Survey \citep[GCLASS;][]{Muzzin2012}. These clusters were initially detected in the \textit{Spitzer} Adaptation of the Red Sequence Cluster survey \citep[SpARCS;][]{Muzzin2009, Wilson2009, Demarco2010}, where they were each located via their overdensity of red-sequence galaxies in shallow $z^{\prime}$ and IRAC $3.6\mu$m images. The VLT detection-band (see sections~\ref{VLTObs} and~\ref{SourceDet}) images of the four clusters studied in this work are shown in Figure~\ref{TCIms}, where we also show the regions in which we initially select cluster members (within 1 Mpc from the cluster centre). We summarise the cluster properties in Table~\ref{Clusterinfo}. 

With all four clusters being part of the widely used GCLASS survey, and one of the four also part of the Gemini Observations of Galaxies in Rich Early ENvironments (GOGREEN) survey \citep{Balogh2017, Balogh2021}, there is a wealth of data available for galaxy cluster science. Accordingly, these clusters have been studied extensively throughout the literature \citep{Lidman2012, Lidman2013, vanderBurg2013, vanderBurg2014, Muzzin2014, Foltz2015, Balogh2016, Biviano2016, Biviano2021, Matharu2019, Matharu2021, Werner2022, Baxter2023}. 

In this work, we combine the already fruitful surveys of GCLASS and GOGREEN with deep VLT observations in filters capable of selecting faint and low-mass cluster members. Details of the VLT observations are given in section~\ref{VLTObs}, with details of the  GCLASS/GOGREEN data given in section~\ref{GCLASSGOGREEN}. 

\subsubsection{VLT observations}
\label{VLTObs}

We obtained deep images of the clusters which sample the rest-frame wavelengths of the Balmer and 4000\,{\AA} breaks of galaxies at the cluster redshifts (see Table~\ref{Clusterinfo}) as part of ESO programme 099.A-0058 (P.I.~Hatch) using HAWK-I \citep{KisslerPatig2008} and FORS2 \citep{Appenzeller1998} on the ESO VLT. For clusters SpARCS~0034 and 0036, located at $z\approx0.87$, we use  the FORS2 filters centred at 691\,nm and 834\,nm. For SpARCS~0215, we use the  FORS2 filters centred at 691\,nm to probe blueward of the breaks, and two filters to probe above the breaks: 834\,nm and the $z_{\rm special+43}$ filter. For SpARCS~0035 we use the FORS2 filter centred at 834\,nm and the HAWK-I broad-band $Y$ filter. Throughout this work we refer to the image blueward of the Balmer and 4000\,{\AA} breaks as $F_b$, and the image redward of the breaks as $F_r$. For SpARCS~0215, which has two filters redward of the break, we chose the image taken through the 834\,nm filter to be $F_r$ since it brackets the breaks more tightly. The [OII]3727 emission line falls within the $F_b$ filter for three of the clusters which may bias the fluxes, but sources affected by this are removed by the red-sequence selection in section~\ref{RSsel}. Two partially overlapping images were taken through each filter  to cover as much as possible of the $10$\,arcmin\,$\times10$\,arcmin field of view as the GLASS/GOGREEN data.

The images were reduced using the publicly available {\sc theli} software \citep{Erban2005, Schirmer2013} following the usual steps of bias correction, flat fielding and background subtraction. The public GCLASS and GOGREEN catalogue (see section\,\ref{GCLASSGOGREEN}) was used as the basis for the astrometric calibration before the individual exposures were combined using {\sc swarp}. Finally, flux calibration was achieved by linearly interpolating the photometry of the GCLASS/GOGREEN catalogue. PSF-homogenised images were created by smoothing the sharper image with a Gaussian kernel until the growth curves of stars within each field were consistent. The FWHM of the PSF-homogenised frames were 0.55, 0.67, 0.75, and 1.1\,arcsec for  SpARCS~0034, 0036, 0215, and 0035, respectively. See Table~\ref{Clusterinfo} for details about the depth and completeness limits of the $F_r$ images.

\subsubsection{The GCLASS and GOGREEN data}
\label{GCLASSGOGREEN}

We used data from the first public data release (DR1) of the GCLASS and GOGREEN surveys that are presented in \cite{Balogh2021}. The data from DR1\footnote{\url{https://www.canfar.net/storage/vault/list/GOGREEN/DR1}} used in this work include the $K_s$-selected multiwavelength photometry catalogues, gaussianised-PSF stacked images, inverse-variance weight maps and bright-star masks. In this work, we create our own $F_r$-selected catalogues and so we perform aperture photometry on the PSF-homogenised stacks ourselves in section~\ref{SourceDet}. We therefore do not use the actual photometry from the DR1 photometry catalogues, other than to measure the zeropoints for each filter via the matching of stars (see section~\ref{ZPs} for details).

The images used in this work were observed in the following filters \citep[see][for more information on these observations]{Lidman2012, Balogh2021}:

\hangpara{0.7cm}{1}SpARCS 0034 - \\ $ugri$ with IMACS on Magellan, $z$-band with MOSAIC-II on the Blanco telescope at CTIO, $J$ and $K_s$ bands with ISPI on the Blanco telescope at CTIO, F140W with the WFC3 on HST, and IRAC channels 1 through 4 on \textit{Spitzer}, making a total of 14 filters when including the 691\,nm and 834\,nm FORS2 observations from section~\ref{VLTObs}.

\hangpara{0.7cm}{1}SpARCS 0035 - \\ $UBVRI$ with VIMOS on the VLT, $z$-band with DECam on the Blanco telescope at CTIO, $JK_s$ with HAWK-I on the VLT, $J1$ with FourStar on Magellan, F140W with the WFC3 on HST, and IRAC channels 1 through 4 on \textit{Spitzer}, making a total of 16 filters when including the and 834\,nm FORS2 and HAWKI-I $Y$ observations from section~\ref{VLTObs}.

\hangpara{0.7cm}{1}SpARCS 0036 - \\ same as SpARCS 0034, making a total of 14 filters when including the 691\,nm and 834\,nm FORS2 observations from section~\ref{VLTObs}.

\hangpara{0.7cm}{1}SpARCS 0215 - \\ same as SpARCS 0034, making a total of 15 filters when including the 691\,nm, 815\,nm and $z_{\rm special+43}$ FORS2 observations from section~\ref{VLTObs}. 

\subsubsection{Source detection and photometry}
\label{SourceDet}

Initial source catalogues were created using \textsc{SExtractor} \citep[SE;][]{Bertin1996}, where detection was performed on the redder of the VLT image pairs for each cluster (i.e. $F_r$). We used a detection threshold of $1.5\sigma$, minimum pixel area per object of 3, minimum contrast ratio of 0.0005, number of thresholds for deblending of 32, and cleaning efficiency of 0.4. Fluxes were measured in $2\arcsec$ diameter apertures using \texttt{photutils} \citep{Bradley2023}, and converted to AB magnitudes using the zeropoints derived in section~\ref{ZPs}. We also converted all aperture fluxes to total fluxes using the [$F_{r,ap}$]/[$F_{r,tot}$] ratio. We calculated this ratio for each galaxy using the fixed-aperture fluxes (\texttt{FLUX\_APER}) and kron-aperture fluxes (\texttt{FLUX\_AUTO}) from \textsc{SExtractor}, measured on the redder VLT image ($F_r$) for each cluster. Prior to the initial detection, we created a global mask which uses a combination of the bright-star masks from the GCLASS/GOGREEN DR1, and the weight maps for the VLT images $F_b$ and $F_r$, and the ($B$)$gzK$ and $J$ images, where any pixel with a coverage of less than $30\%$ of the maximum for a given image is masked. The requirement of significant coverage in the two VLT images are so that we can perform the red-sequence selection (section~\ref{RSsel}), while the requirement in the other 4 filters are so that we can remove star contaminants (section~\ref{StarGalSep}).

Uncertainties on fluxes ($\sigma_f$) were calculated by combining the propagated poissonian error on electron counts ($\sigma_{e}$) with the image depth ($\sigma_{depth}$), in quadrature (i.e. $\sigma_f = \sqrt{\sigma_{e}^2 + \sigma_{depth}^2}$). The propagated Poissonian error on electron counts is simply $\sigma_{e} = \textrm{\textit{ADUs}}\times\sqrt{N_e} / N_e$, where the number of electrons $N_e = \textrm{\textit{ADUs}}\times \textrm{\textit{exposure time}} \times \textrm{\textit{gain}} $. The image depth, or limiting flux, was calculated using \texttt{photutils}, by calculating the standard deviation of fluxes in apertures placed on random blank regions of the image. The $5\sigma_{depth}$ limits shown in Table~\ref{Clusterinfo} have been converted to AB magnitudes using the corresponding zeropoints.

\subsubsection{Photometric calibration}
\label{ZPs}

As explained in \cite{Balogh2021}, the photometric zeropoints for the $J$ and $K$ filters were calibrated with respect to 2MASS \citep{Jarrett2000}, while the calibration of the other GCLASS/GOGREEN DR1 filters was based on the universality of the stellar locus \citep{High2009}. 

Upon measuring fluxes in ADUs in the images, we matched a sample of stars in our catalogue with the photometry catalogue from DR1. We extracted this sample of stars using a combination of criteria detailed in section~\ref{StarGalSep}, as well as a visual inspection of the F140W HST images, where stars have diffraction spikes and can therefore be easily located. However, rather than using the disjunctive logic from section~\ref{StarGalSep} to remove sources that could possibly be stars, we instead used conjunctive logic to select a robust sample of stars\footnote{The criteria used was \texttt{CLASS\_STAR} $> 0.8$ $\land$ Eq.~\ref{gzkeq} $\land$ Eq.~\ref{JKeq}, with additional stars coming from the visual inspection (see section~\ref{StarGalSep} for more details)}. Using this sample of stars ($\sim60$ per cluster), we compared the fluxes we measured to the fluxes in the DR1 photometry catalogues, measuring the mean shift as the zeropoint offset for a given filter. 

\subsubsection{PSF-homogenisation}
\label{PSFhom}

As explained in section~\ref{VLTObs}, the $F_b$ and $F_r$ images are homogenised to each other. This section describes the homogenisation of \textit{all} images (given in section~\ref{GCLASSGOGREEN}) to make them ready for the SED fitting performed in section~\ref{RFcol}. 

In order to produce an accurate and representative measurement of galaxy properties, the observed photometry must come from the same intrinsic part of the galaxy. However, with the photometry derived from a variety of telescopes/instruments, the varying PSF from each observation will result in different parts of the galaxy being represented in the photometry. To combat this, we degraded the PSF from all observations from a given cluster to match the observation with the largest PSF.

To measure the PSF for each observation, we used the same sample of stars that are used to derive zeropoints in section~\ref{ZPs}. For each star, we measured fluxes in 200 concentric apertures with radii $0\arcsec < r \leq 5\arcsec$. We converted these fluxes into a surface brightness within an annulus, by subtracting the flux in a given aperture from that in the subsequent aperture and dividing by the area of that annulus. Plotting this against the annuli radius gives us the shape of the PSF. We fit a Gaussian to the radial profile, outputting the full width at half maximum (FWHM). We estimated the FWHM of each observation as the mean of the FWHMs measured for each of the stars, making sure to remove any that have nearby bright sources. We degraded the PSFs of all observations through a convolution with a Gaussian kernel, until the FWHMs match the observation with the largest PSF to within $5\%$. This results in PSF sizes of $1.56\pm0.04\arcsec$ for SpARCS 0034, $2.08\pm0.05\arcsec$ for SpARCS 0035, $1.97\pm0.05\arcsec$ for SpARCS 0036, and $1.98\pm0.05\arcsec$ for SpARCS 0215. 

The average PSF FWHM for the IRAC images across the four clusters is $2.77\arcsec$, which is significantly larger than the superior spatial information from the ground-based imaging. The IRAC observations were therefore not included in the PSF homogenisation. Previous studies overcame this through the measurement of fluxes in different sized apertures \citep[see][]{Quadri2007}, though we verified that the rest-frame colours and stellar masses output from \texttt{EAZY} are negligibly affected whether or not we include the IRAC photometry.

\subsection{Field sample}
\label{FieldSample}

To determine the effects of environmental quenching in clusters, we need a sample of passive galaxies from an environment that is less affected by environmental quenching, where the only (or at least dominant) quenching mechanisms are internal, i.e. a field sample. The field sample used in this work needs to be extracted from a survey that satisfies requirements in four key areas; depth, area, redshifts and photometric bands. The data need to be deep enough to probe the lower masses (at least to the level we reach in the clusters), wide enough to have good statistics, redshifts accurate enough to select galaxies at the cluster redshift (at least within $\pm0.025$), and use the same or similar filters that have been used for the clusters so that we can reproduce the method that has been applied to the clusters on the field. Unfortunately, no single survey exists that meets all the requirements. Therefore, we use a combination of two surveys: COSMOS2020 and PAUS.

\subsubsection{COSMOS2020 survey}

The latest release of the COSMOS catalogue is COSMOS2020 \citep{Weaver2022}. It contains over 1 million sources selected in a \textit{izYJHK$_s$} coadd, reaching depths down to 26 mag. This data is comprised of photometry ranging from UV to 8 $\mu$m over the total 2 deg$^2$ area, provided by a combination of surveys and instruments. These are: $YJHK_s$ from the latest release of the UltraVISTA survey \citep[DR4;][]{McCracken2012, Moneti2023}, $grizy$ from Subaru's Hyper Suprime-Cam instrument \citep[HSC PDR2;][]{Aihara2019}, $u$ from the CLAUDS survey taken by the Canada-France-Hawaii Telescope \citep{Sawicki2019}, \emph{Spitzer}/IRAC channels one through four \citep{Euclid2022}, NB118 from VISTA \citep{MilvangJensen2013}, and 7 broad-bands ($B$, $g^{+}$, $V$, $r^{+}$, $i^{+}$, $z^{+}$, $z^{++}$), 12 intermediate bands (\textit{IB427, IB464, IA484, IB505, IA527, IB574, IA624, IA679, IB709, IA738, IA767, IB827}) and two narrowbands (\textit{NB711, NB816}) from Suburu/Suprime-Cam \citep{Taniguchi2007, Taniguchi2015}. The use of over 30 photometric bands in the redshift estimation allows for sub-percent accuracy \citep{Weaver2022}.

It is clear that the COSMOS2020 survey meets the depth, area and redshift requirements for the field sample. However, the closest bands they have available to the ones we use (FORS2 bands 691 nm, 815 nm, 834 nm and HAWK-I $Y$) are the Subaru/Suprime-Cam bands \textit{IA679, NB816, IB827} and VISTA $Y$, respectively. These are similar, but are offset by a few nm for NB691 and NB834, as shown in Figure~\ref{photbands}. This offset means that we cannot accurately reproduce the method we apply to the clusters to a field sample extracted from COSMOS2020 alone.

\subsubsection{PAU survey}

\begin{figure}   \includegraphics[width=\columnwidth]{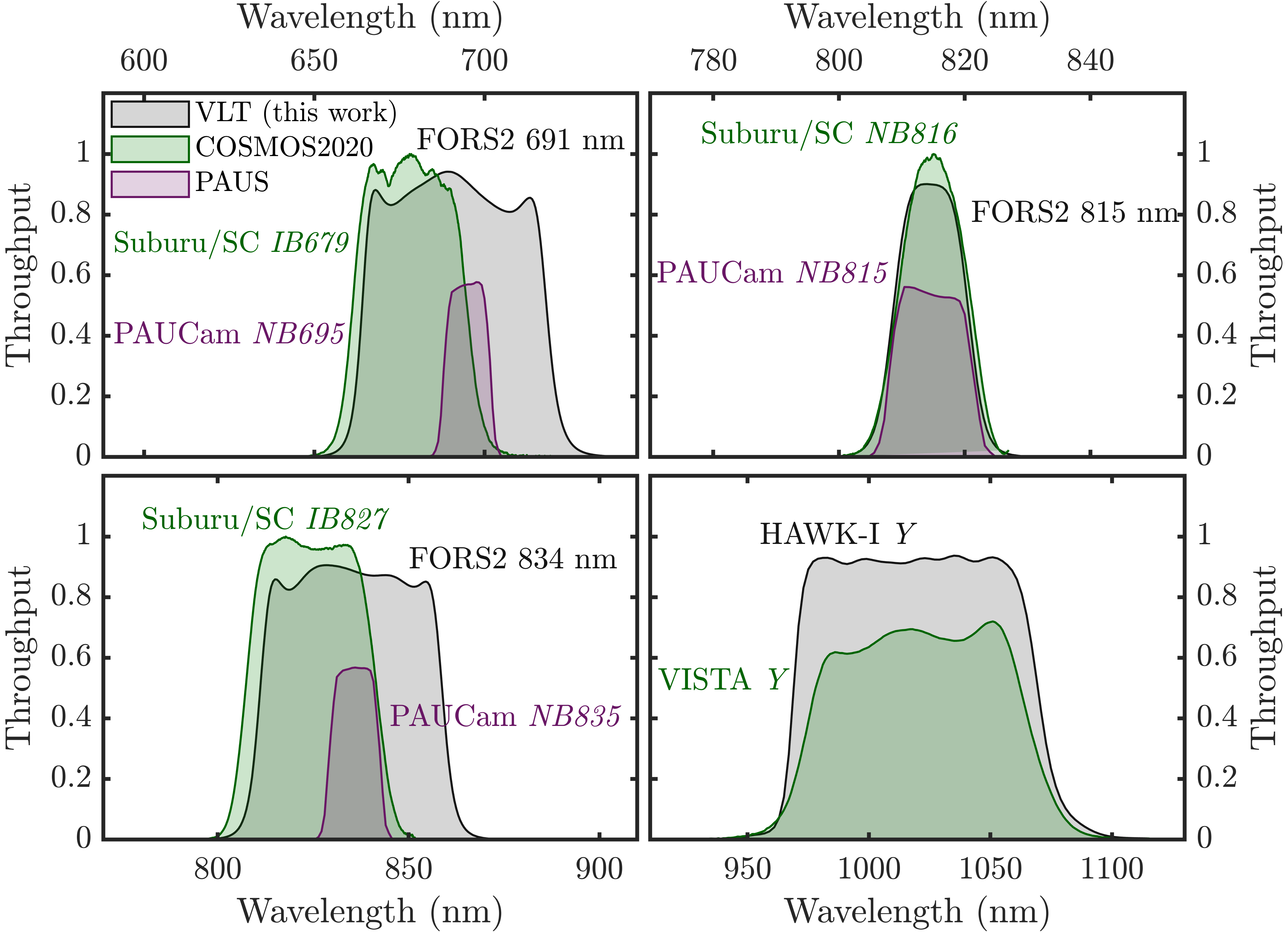}
    \caption{\small{Transmission curves of the $F_b$ and $F_r$ filters used in this work (grey), with the nearest available filters in COSMOS2020 (green) and PAUS (purple).}}
        \label{photbands}
\end{figure}

The Physics of the Accelerating Universe survey \citep[PAUS;][]{Benitez2009} is a narrow-band photometric survey taken at the William Herschel Telescope, using the purpose-built PAU Camera \citep{Padilla2019}. The PAU Camera is fitted with 40 narrow-band filters uniformly spaced between 455 nm and 845 nm in steps of 10 nm, which enables photometric redshift measurements with sub-percent accuracy. In total, the PAU survey covers $\sim50$ deg$^2$ across five fields, and is complete up to $i \leq 23$ AB. The closest bands available to the ones used in this work are \textit{NB690, NB815}, and \textit{NB835}. As shown in Figure~\ref{photbands}, the central wavelengths of these bands are closer to the bands we use in this work than the COSMOS2020 bands, with a maximum offset of only 1 nm.

The PAU survey meets our field sample requirements for the area, redshifts and photometric bands, but only reaches a depth of $i \leq 23$. This means we cannot probe down to the same low masses we can in the clusters using the PAU survey alone.

\subsubsection{Combining COSMOS2020 and PAUS}

With neither the COSMOS2020 or PAU surveys satisfying our field sample requirements alone, we use the two in conjunction. As one of the PAU survey winter fields is COSMOS, we are able to match sources that are in both surveys. This, by definition, will only contain the brightest sources in the COSMOS2020 catalogue (with PAUS only reaching a depth of $i \leq 23$). But, for these sources, we have a measurement of magnitude in both the COSMOS2020 filters and the PAUS filters. We compared the magnitudes that are measured in the COSMOS2020 filters (that are slightly offset to the VLT filters) to those that are measured in the PAUS filters (that we assume have consistent central wavelengths with the VLT filters). We fit a linear relationship between the magnitudes measured in the COSMOS2020 and PAUS filters, which was extrapolated to the fainter magnitudes not reached in PAUS, to shift all the magnitudes in the COSMOS2020 catalogue. At the brighter end, these shifts are negligible for $F_b$ and $F_r$ in all four clusters. At the fainter end, the shift is larger, at a maximum of 0.4 mag for a galaxy with magnitude 25. While not ideal, these slight shifts allow us to create a more representative field sample, and allow for a more accurate comparison with our cluster sample. As the VISTA $Y$ and HAWK-I $Y$ bands are so similar, we do not need to perform any adjustment like we do with the other filters.

With the magnitudes adjusted, we selected our initial field samples as galaxies with a redshift $\pm0.025$ from the corresponding cluster redshift. This range corresponds to the typical photometric redshift uncertainty for galaxies in the field samples.  Other than the statistical background subtraction (section~\ref{SBS}), all selection criteria performed on the cluster samples are also performed on the field samples.

We also note here that significant overdensities have been detected in the COSMOS field at various redshifts \citep{Kovac2010, Laigle2016, Wang2016, Darvish2017, Darvish2020, Cucciati2018, Gozaliasl2019, Forrest2023}. As these structures mostly consist of groups, filaments and high-redshift protoclusters, we expect the dominant quenching mechanisms in the field to be internal. Though, some signatures of environmental quenching will remain in the analysis of the field samples. We discuss this further in section~\ref{Disc}.

\section{Analysis}
\label{Analysis}

\subsection{Star-galaxy separation}
\label{StarGalSep}

\begin{figure*}   \includegraphics[width=\textwidth]{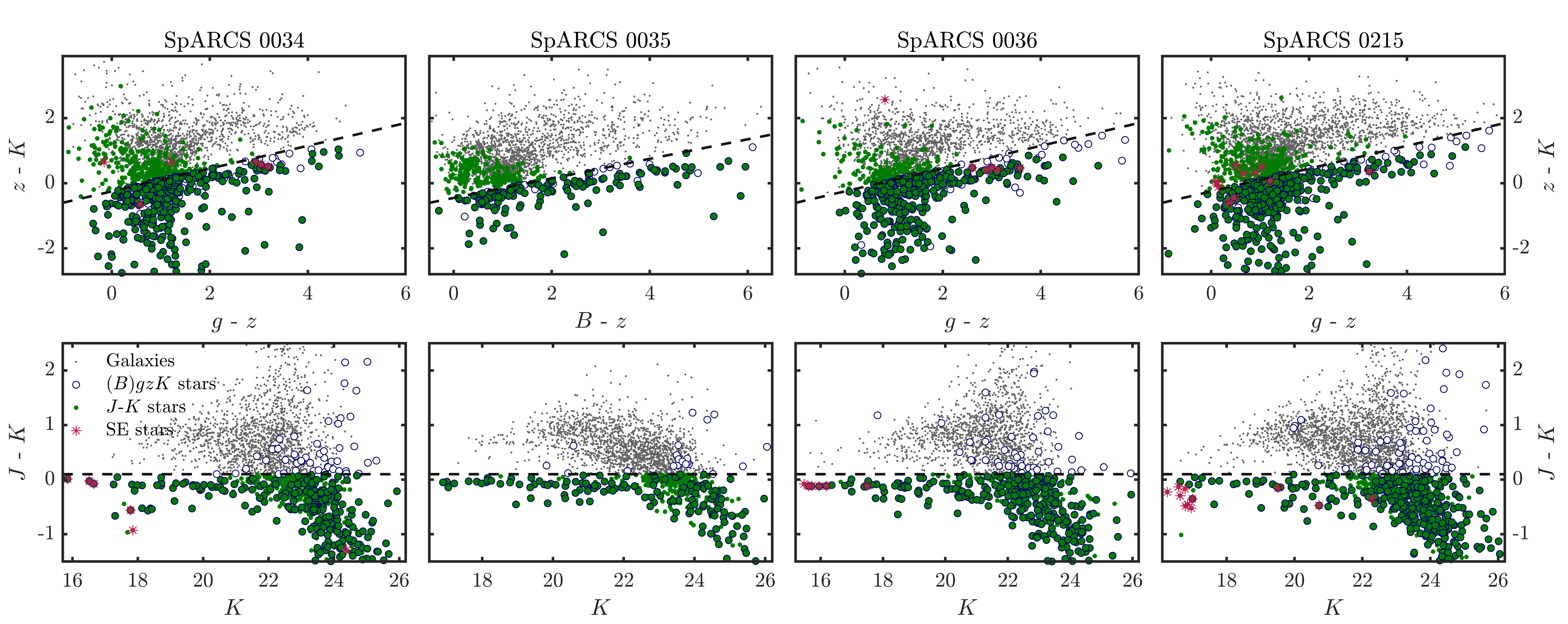}
    \caption{\small{\textit{Top:} $(B)gzK$ colour-colour plots for all sources detected in the $F_r$ image. The equations of the black dashed lines are defined by equation~\ref{gzkeq}, where any source below one of these lines (shown in blue circles) is classified as a star. \textit{Bottom:} $J-K$ CMD for all sources as in the row above. The equation of the black dashed lines indicates our stellar locus cut of $J-K < 0.1$ (equation~\ref{JKeq}), where any source below one of the lines (green dots) is classified as a star. For both rows, the pink asterisks show the positions of the stars classified by \textsc{SExtractor}. Combinations of symbols show the sources classified as stars through multiple methods. The grey points are sources that are not classified as a star via any method, and so are considered galaxies.}}
        \label{gzkplots}
\end{figure*}

To distinguish between galaxies and foreground stars, we used a combination of morphological and colour criteria. SE provides the \texttt{CLASS\_STAR} output parameter which indicates the likelihood of a source being a star or galaxy using a neural network that is trained on simulated images. This morphological classification works on the assumption that galaxies appear `fuzzier' than stars due to their extended nature. We classify SE stars as sources with \texttt{CLASS\_STAR} $> 0.8$ in the $F_r$ detection image, which are shown as pink asterisks in Figure~\ref{gzkplots}. This type of morphological classification breaks down at fainter magnitudes, where SE randomly assigns a stellarity value between 0 and 1. Therefore, we also utilise the $BzK_s$ criteria \citep{Daddi2004} and its $gzK$ adaptation \citep[e.g][]{Arcila-osejo2013,Ishikawa2015}, to isolate the stellar locus. The $BzK_s$ technique is commonly used to distinguish between star-forming and passive galaxies at $1.4<z<2.5$, but is also able to separate stars and galaxies. By a visual inspection of the $(B)gzK$ diagrams shown in the top row of Figure~\ref{gzkplots}, we identified the stellar locus using

\begin{align}
\label{gzkeq}
    &(z-K_s) - 0.3(B-z) < -0.45\, \textrm{ and} \\
    &(z-K) - 0.35(g-z) < -0.25 , \nonumber
\end{align}
\\
\noindent where stars identified via this method are highlighted with blue circles. Additionally, we used a near-infrared colour criterion as a galaxy-star discriminant. By a visual inspection of the colour-magnitude diagrams (CMDs) in the bottom row of Figure~\ref{gzkplots}, we identified the stellar locus using

\begin{equation}
\label{JKeq}
    (J-K) < 0.1 ,
\end{equation}
\\
\noindent where stars identified via this method are highlighted with green dots. Stars that have been identified with more than one of the star classification methods outlined above are highlighted with the corresponding combination of symbols/colours. If a source meets any of these criteria, it is removed from the analysis. This is a conservative approach, which will lead to the removal of a small number of galaxies. The slight undercount in galaxies is outweighed by the gain in sample purity, leading to more reliable SMF measurements.

\subsection{Red-sequence selection}
\label{RSsel}

The red-sequence feature of colour-magnitude diagrams is commonly used to separate red, passive galaxies from bluer star-forming galaxies. The slope, scatter and location of this feature are able to place constraints on the formation epoch of stars in the passive galaxies \citep{Bower1992, Bower1998, Ellis1997, Kodama1998}. As galaxy clusters are rich with passive galaxies, they are the environments in which the red-sequence is most conspicuous, even up to redshifts of $z\sim2$ \citep[e.g.][]{Tanaka2010, Gobat2011, Spitler2012, Stanford2012, Andreon2014}, and possibly beyond \citep[e.g.][]{McConachie2022, Ito2023, Tanaka2024}. Without spectroscopic redshifts for the faint, passive galaxies, we can either attempt to select cluster galaxies using photometric redshifts, or via a redshift-independent method. \cite{vanderBurg2013} select cluster members using both spectroscopic and photometric redshifts. For galaxies without a $z_{spec}$, they use the fractions of false positives and false negatives to correct the number counts for cluster membership. Typically, the galaxies that do not have $z_{spec}$ measurements (i.e. just a $z_{phot}$) are the faint/low-mass ones. We would therefore be unable to perform an accurate correction for the faint/low-mass galaxies. Therefore, with the large uncertainties associated with photometric redshifts for the faint galaxies which are not detected in all of the broad-band filters, we opt for the redshift-independent method that relies on the optical/near infrared colours of [$F_b$] - [$F_r$]. This red-sequence selection enables us to accurately select galaxies whose 4000\,{\AA} breaks have been redshifted to within the wavelengths of the two VLT $F_b$ and $F_r$ filters, thereby removing galaxies not at the redshift of the given cluster. The 4000\,{\AA} break is mostly caused by the absorption of several ionised metallic elements, with a contribution from the latter lines in the Balmer series. Therefore, selecting galaxies based on this spectral feature will preferentially select passive galaxies, which have stronger 4000\,{\AA} breaks. We still expect there to be significant contribution from star forming galaxies upon making the red-sequence selection as these galaxies can appear red due to dust, and so we also perform a passive/star-forming separation based on rest-frame colours (details in section~\ref{RFcol}).

\begin{figure*}   \includegraphics[width=\textwidth]{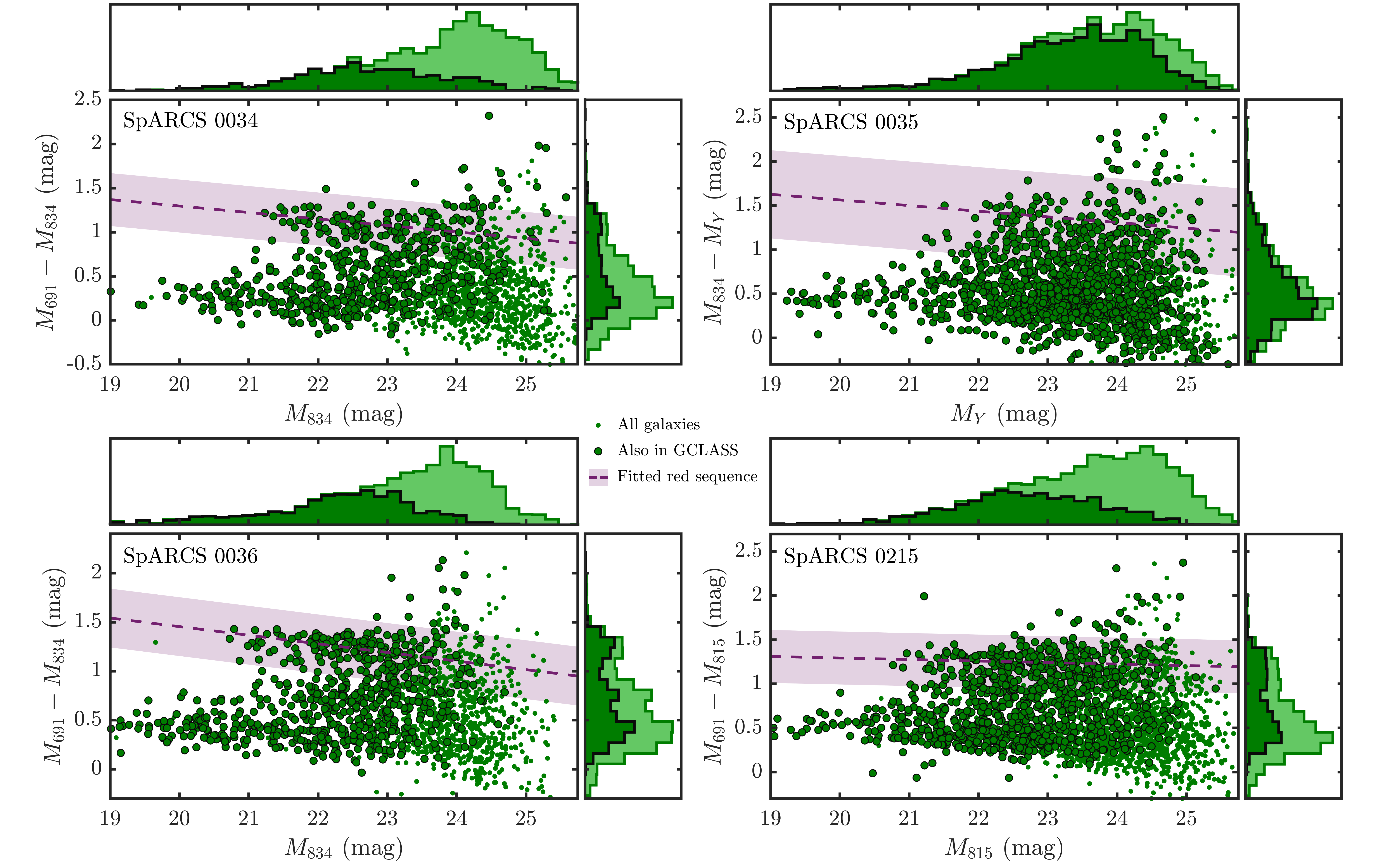}
    \caption{\small{Colour-magnitude diagrams for each of the clusters. The points shown are all the sources remaining after the star removal, with those that are also in GCLASS/GOGREEN outlined with black circles. The distributions above and to the side represent the magnitude and colour distributions of galaxies in GCLASS/GOGREEN (black outline) and the new sources detected in section~\ref{SourceDet} (light-green and green outline), respectively. The dashed purple lines show the fitted red sequence for each cluster, with the shaded region showing $\pm0.3$ mag (or $\pm0.5$ mag for SpARCS 0035) around the fitted red sequence, which is used to select galaxies on the red sequence.}}
        \label{CMDs}
\end{figure*}

The CMDs of the four clusters are shown in Figure~\ref{CMDs}, along with the fitted observed red sequence. The red sequence was fit using linear regression that is weighted to the inverse of the magnitude uncertainties to a pre-selected sample that is $a)$ within the central region ($r < 1$ Mpc) of the respective cluster and $b)$ in the red locus of galaxies in the CMD (which is visually identified). We find a mean red-sequence slope across the four clusters of $-0.06\pm0.01$, which is similar to the slope used in \citeauthor{Chan2019} (\citeyear{Chan2019}; and references therein). The red-sequence galaxies were selected as those $\pm0.3$ mag around the robust fit, except for SpARCS 0035 in which we selected red sequence galaxies in a larger range of $\pm0.5$ mag. The larger selection region for SpARCS 0035 is due to its indistinct red sequence. We verified that our main conclusions do not change if we alter this selection to galaxies $\pm0.5$ mag (or $\pm0.7$ mag for SpARCS 0035) from the red-sequence fit. In each CMD we highlight the additional galaxies which are too faint to be included in the \cite{vanderBurg2013} study (i.e. those without black outlines).

\begin{figure*}   \includegraphics[width=\textwidth]{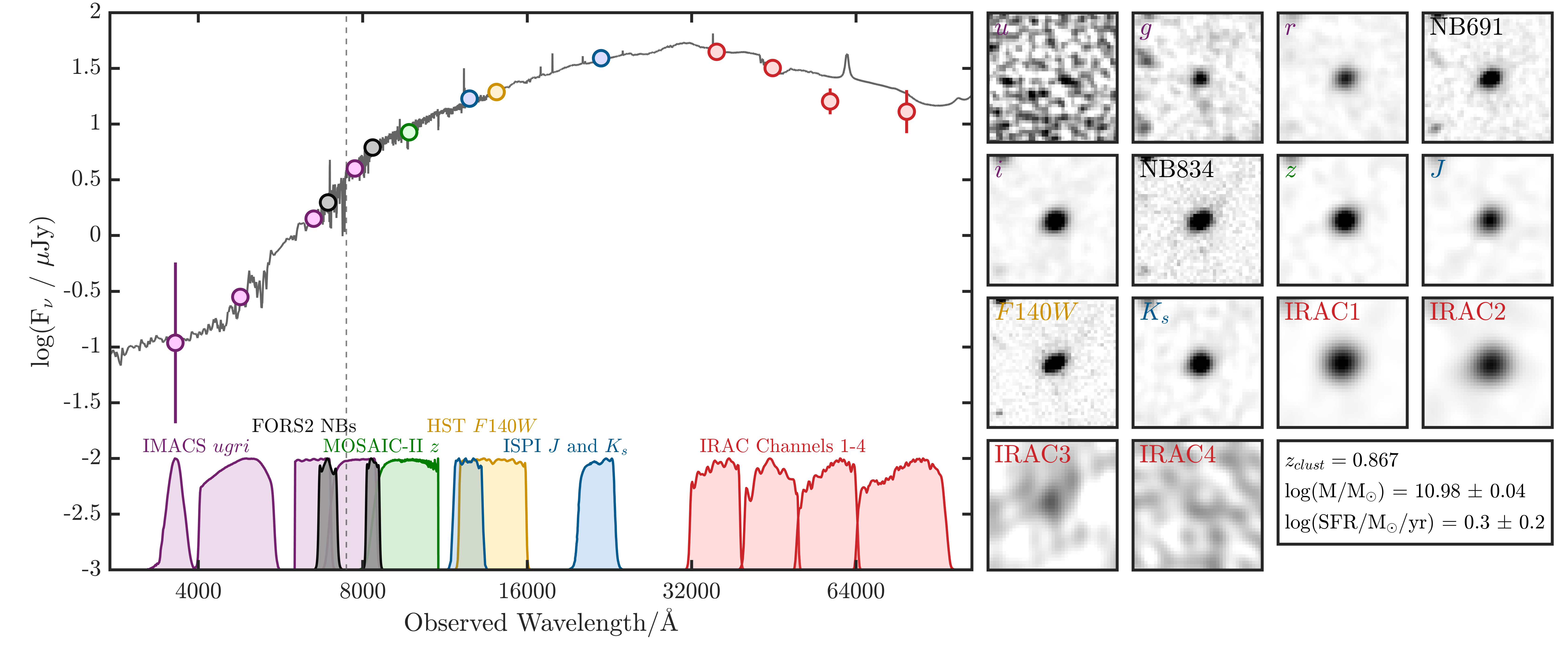}
    \caption{\small{\textit{Left}: Example of an \texttt{EAZY} Spectral Energy Distribution (grey) fit to the photometry (coloured circles) of a galaxy in SpARCS 0034 of mass $9.55\times10^{10}$ M$_{\odot}$. Also shown are the transmission curves of the filters used to measure the photometry for galaxies in this cluster. \textit{Right}: $10\arcsec\times10\arcsec$ cutouts of the PSF-homogenised images used to measure the photometry shown in the left panel, centered on the same galaxy. In all panels, the colours distinguish between the instruments used to measure the photometry.}}
        \label{SEDfit}
\end{figure*}

\subsection{Galaxy properties}
\label{RFcol}

To measure the rest-frame colours and stellar masses of the selected galaxies, we used the Python version of the template-fitting code \texttt{EAZY}\footnote{\url{https://www.github.com/gbrammer/eazy-py}} \citep{Brammer2008}. When running \texttt{EAZY}, we used the same set of 17 templates derived from the Flexible Stellar Population Synthesis models \citep[FSPS;][]{Conroy2009, Conroy2010} that are used in the creation of the COSMOS2020 catalogue, which have a variety of dust attenuation and ages from log-normal star formation histories that broadly span the rest-frame $UVJ$ colour-space of $0 < z < 3$ galaxies. These templates were fit to the observed photometry in a non-negative linear combination, as shown in Figure~\ref{SEDfit}, where we fixed the redshift of all galaxies to the cluster redshift.

In addition to the red-sequence selection which preferentially selects passive galaxies, we also used rest-frame $U-V$ and $V-J$ colours to remove star-forming galaxies. These colours are effective in discriminating between passive and star-forming galaxies\footnote{If redshifts are not known to a high accuracy, it is possible for significant degeneracies between redshift, dust, age, and metallicity to be introduced. Even so, the affects of contaminants resulting from this are removed via the background subtraction in section~\ref{SBS}.}, and are even impervious to dust-reddening \citep[e.g.][]{Wuyts2007,Williams2009,Patel2012}. The criteria we adopted to select passive galaxies are given by

\begin{equation}
\begin{split}
    (U-V) &> 0.88 \times (V-J) + 0.59 \qquad\quad [0.5 < z < 1.0]\\
    (U-V) &> 0.88 \times (V-J) + 0.49 \qquad\quad [1.0 < z < 1.5] ,
\end{split}
\label{uvjcrit}
\end{equation}
\\
with additional criteria of $U-V > 1.3$ and $V-J < 1.6$ to remove unobscured and dusty star-forming contaminants.

To measure the rest-frame $UVJ$ colours of galaxies, we used \texttt{EAZY}, fixing the redshifts of all the red sequence-selected galaxies to the cluster's redshift. The fluxes we input into \texttt{EAZY} were measured on PSF-homogenised images (see section~\ref{PSFhom}) in $2^{\prime\prime}$ diameter apertures placed on the same locations as the sources in the $F_r$-selected catalogues from section~\ref{SourceDet}. The flux errors we input into \texttt{EAZY} are simply $1\sigma$ background estimates for the respective images, as the uncertainties for fainter objects are dominated by the background as opposed to electron counts.

We encountered slight offsets between the $UVJ$ colour distributions measured in each of the clusters and that in the COSMOS2020 field. These offsets have been found in numerous previous studies using similar cluster samples \citep[e.g.][]{Whitaker2011, Muzzin2013, Skelton2014, LeeBrown2017, vanderBurg2018, vanderBurg2020}, and are indicative of residual uncertainties in the initial photometric calibration. Therefore, we manually applied shifts to the $U-V$ and $V-J$ colours of the cluster galaxies by aligning their quiescent loci to the quiescent loci in the respective COSMOS2020 field sample, as defined in Section~\ref{FieldSample}. 

\begin{figure*}   \includegraphics[width=\textwidth]{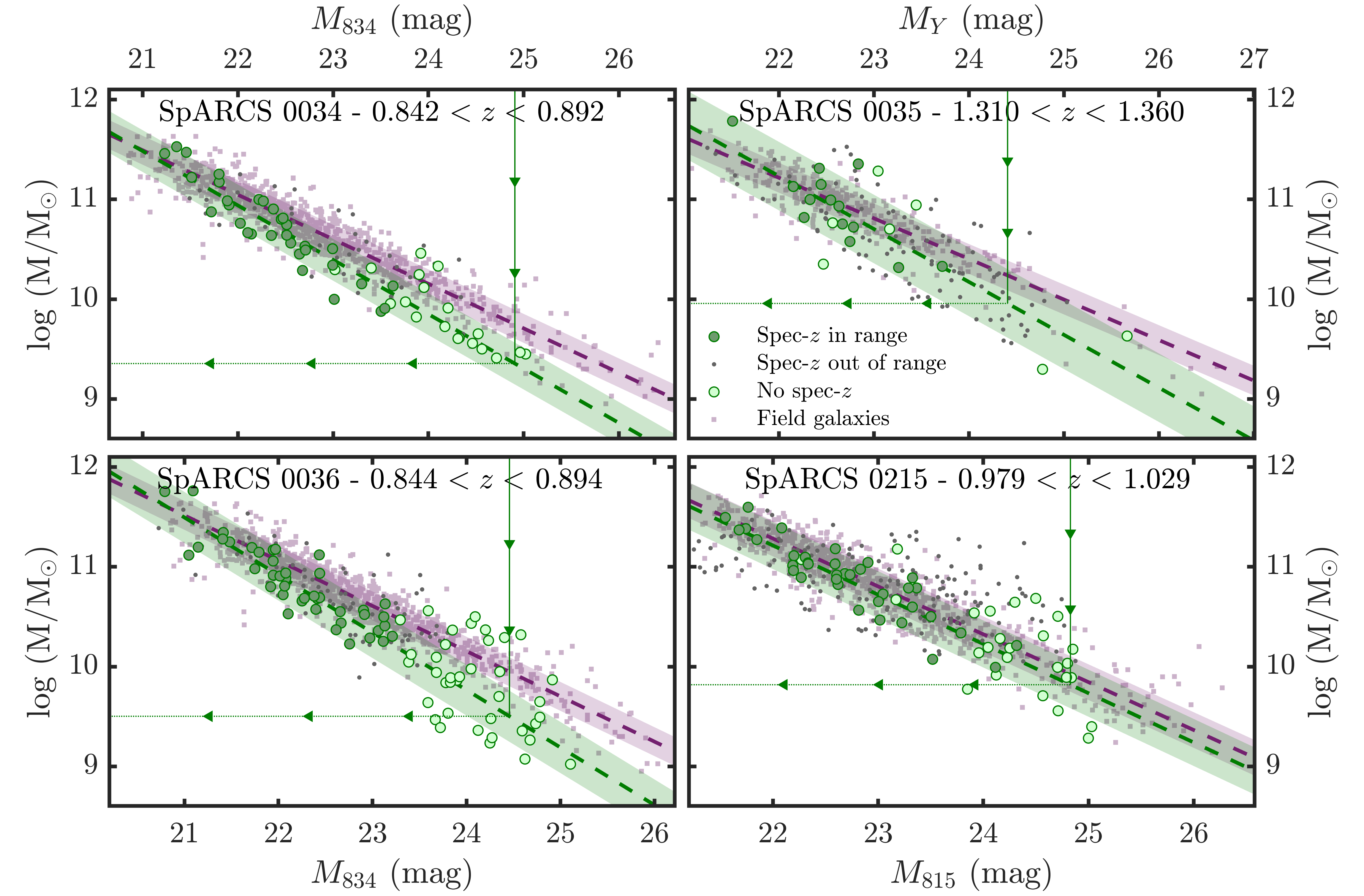}
    \caption{\small{The relationship between stellar mass and magnitude for passive galaxies selected in the red sequence of each of the clusters (green), and the corresponding field sample (purple). The darker green points represent galaxies that have a spec-$z$ within the same range used to select the field sample (given at the top of each panel), with the lighter green points showing those without a spec-$z$ measurement (i.e. new sources detected in the $F_r$ images). The green and purple dashed lines show a fit to the mass-luminosity relation for the cluster and field points using a robust linear regression, with 1$\sigma$ errors shown by the shaded regions. The vertical and horizontal lines show the magnitude and mass limits used to limit data fit with a Schechter function in section~\ref{Schfits}. The direction of arrow and line style determine whether the limit was explicitly measured (solid) or implicitly derived (dotted). For example, the magnitude limit for the clusters is explicitly measured via an injection-recovery simulation (section~\ref{compcorr}), whereas the mass limit for clusters is implicitly derived using the magnitude limit and best fit of the mass-luminosity relation. The grey dots show galaxies with spec-$z$s outside of the redshift range given, showing the need for a statistical background subtraction (section~\ref{SBS}) to remove contaminants.}}
        \label{MassMag}
\end{figure*}

Stellar masses were also estimated using \texttt{EAZY}, where the redshifts were fixed to the redshift of the cluster. The FSPS templates that are fit to the measured photometry were created assuming a \cite{Chabrier2003} initial mass function, a \cite{Kriek2013} dust law, and solar metallicity\footnote{This assumption can under-estimate stellar mass measurements for low-mass galaxies by $\sim0.25$ dex \citep{Bellstedt2024}. This may impact the shape of the SMFs, but it is consistent with previous works \citep[e.g.][]{vanderBurg2013}, and does not affect our cluster versus field galaxy comparisons.}. The relationship between stellar mass and magnitude is shown in Figure~\ref{MassMag}. We fit this relation using a robust linear regression, which is used to estimate the mass completeness from the magnitude completeness that is measured in section~\ref{compcorr}. For galaxies with log(M/M${_\odot}$) $ > 10$, the typical uncertainty on the mass estimate is $\sim0.05$. For galaxies with log(M/M${_\odot}$) $ < 10$, the typical uncertainty on the mass estimate is $\sim0.2$ dex. 

Having performed all the selection criteria, our cluster samples consist of 52, 55, 81, and 95 galaxies (this includes selecting those within 1 Mpc from the cluster centre). The corresponding field samples consist of 896, 300, 919 and 756 galaxies in clusters SpARCS 0034, 0035, 0036 and 0215, respectively. All of these samples include interloper contamination.

\subsection{Kernel density estimation}
\label{KDEM}

Numerous methods have been devised to measure LFs and SMFs \citep[see][for a review]{Johnston2011}, of which the most popular is the binned method. This method was first introduced as the $1/V_{max}$ estimator \citep[e.g.][]{Schmidt1968, RowanRobinson1968}, and has remained widely used in the literature ever since. The main drawback of binned methods is the seemingly arbitrary choice of the bin centre and width, which can dramatically affect the shape of the LFs and SMFs. This issue is particularly prevalent when dealing with low numbers of galaxies. As a result, the shape of the parametric form (section~\ref{Schfits}) that is fit to the binned points is significantly affected. These issues are usually dealt with by combining large samples of clusters to avoid low number densities. As we are attempting to measure individual cluster LFs and SMFs, we opt for a different method entirely.

Kernel Density Estimation \citep[KDE;][]{Rosenblatt1956, Parzen1962} avoids binning data altogether. It works by allowing each galaxy to contribute a smooth Gaussian-shaped `bump' to the LF/SMF. These bumps are summed over to obtain a probability density function that can be normalised to the units of a typical LF/SMF. If we let $\mathbf{M} = (M_1, M_2, \ldots, M_n)$ represent the data points, then the measurement of the LF/SMF will be

\begin{equation} 
\label{KDEeq}
\hat{\varphi}(\mathcal{M}) = \frac{n}{\textrm{Area}} \cdot \frac{1}{nh\sqrt{2\pi}} \sum_{i=1}^{n} e^{-\frac{(\mathcal{M}-M_i)^2}{2h^2}} ,
\end{equation}
\\
where $n$ is the total number of galaxies, $h$ is the bandwidth of the Gaussian kernel, $\mathcal{M}$ are the mass/magnitudes at which the KDE is evaluated, $M_i \in \mathbf{M}$ are the individual data points, and the first term normalises the probability density function into the correct units. The bandwidth was determined using the \cite{Silverman1986} rule-of-thumb estimator $h=0.9\,\textrm{min}\left(\sigma, \frac{IQR}{1.34}\right)$, where $\sigma$ and $IQR$ represent the standard deviation and interquartile range of the data $\mathbf{M}$, respectively. So long as a reasonable number of evaluation points are chosen, the shape of $\hat{\varphi}(\mathcal{M})$ will remain unchanged regardless of the points chosen. However, choosing too many points will artificially reduce the uncertainties associated with the Markov Chain Monte Carlo (MCMC) fits described in section~\ref{Schfits}. Therefore, we evaluated $\hat{\varphi}(\mathcal{M})$ at uniformly distributed points between the minimum and maximum of $\mathbf{M}$, with spacing equal to the bandwidth $h$.

One issue associated with KDE is the boundary bias problem \citep[e.g.][]{Muller1999, Yuan2020}. This problem occurs when data points near a boundary (e.g. magnitude or mass limit) have kernels that extend beyond the boundary, causing them to lose their full probability weight. Also, points that lie just outside the boundary have no contribution to the final probability density function at all. This leads to an underestimation of density near the boundary. To mitigate this, we allowed $\mathbf{M}$ to include galaxies fainter or less massive than the completeness limits derived in section~\ref{compcorr}. This partially alleviates the boundary bias problem by allowing the full probability weights and contribution of all points into $\hat{\varphi}(\mathcal{M})$, though it is still limited to the extent that faint galaxies are within the sample to begin with. Once $\hat{\varphi}(\mathcal{M})$ was computed using the full dataset, we restricted the Schechter function fit (section~\ref{Schfits}) to the reliable data region that is above the completeness limit.

To estimate the uncertainties on $\hat{\varphi}(\mathcal{M})$, we used a combination of Monte Carlo simulations and bootstrapping. First, the data points were randomly varied within their measurement uncertainties, assuming Gaussian distributions for the errors. For each Monte Carlo realisation, we performed 100 bootstrap resamples with replacement and computed $\hat{\varphi}(\mathcal{M})$ for each resample. This process was repeated for 100 Monte Carlo realisations, resulting in 10,000 $\hat{\varphi}(\mathcal{M})$ measurements in total. The final uncertainties, $\hat{\sigma}$, were estimated by calculating the standard deviation of $\hat{\varphi}(\mathcal{M})$ across all 10,000 measurements, with the final $\hat{\varphi}(\mathcal{M})$ taken as the mean.

\subsection{Completeness correction}
\label{compcorr}

In order to characterise the completeness of the sources detected in the respective detection images ($F_r$), we performed an injection-recovery simulation. We created mock galaxy stamps using the image simulation tool \textsc{GalSim} \citep{Rowe2015}, which have exponential (i.e. S\'ersic index $n= 1$) light profiles with half-light radii uniformly distributed in the range $1-3$ kpc, ellipticities uniformly distributed in the range $0-0.5$, and magnitudes uniformly distributed between 18 and 26 mag. A total of 40,000 galaxy stamps were created per cluster, with 250 injected into the original detection images at a time (so not to affect the overall properties of the images). These simulated galaxies were placed on random locations throughout the detection images, so long as they do not overlap with real sources in the image or are on low coverage regions. 

We ran the exact same detection algorithm as the main analysis (Section~\ref{SourceDet}) on the images with injected sources, utilising the \textsc{SExtractor} wrapper for Python - \texttt{sewpy}\footnote{\url{https://github.com/megalut/sewpy}}. We measured the completeness as a function of the recovered magnitude, with the magnitudes at which $70\%$ of the sources are still detected reported in Table~\ref{Clusterinfo}. This was also converted into a mass completeness using the relation between mass and magnitude derived in section~\ref{RFcol}. To correct for the completeness, we divided $\hat{\varphi}(\mathcal{M})$ by the completeness at $\mathcal{M}$. 

For a fair comparison, we used the same magnitude and mass limits for the field as we do for the corresponding clusters. As we perform a completeness correction for the clusters, we require a completeness of $\sim100\%$ for the field. \cite{Weaver2022, Weaver2023} measure a mass completeness limit following the method of \cite{Pozzetti2010}. \cite{Weaver2023} present this estimate for a more secure sample than \cite{Weaver2022}, and also provide separate measurements for the star-forming and quiescent populations. The $70\%$ mass completeness limit of quiescent galaxies from \cite{Weaver2023} is given by

\begin{equation} 
\label{Mcompeq}
M_{70} = -3.79\times10^{8}(1+z) + 2.98\times10^{8}(1+z)^2 .
\end{equation}
\\
As this does not give the completeness as a function of mass, we cannot correct for it. However, we find that the completeness limits we derived for the clusters (and used for the field) are on average 1 dex higher than the $70\%$ completeness limits calculated using equation~\ref{Mcompeq}. With our completeness limit an order of magnitude above the field's $70\%$ completeness limits, we can be confident that the field sample is, to all intents and purposes, complete. 

\subsection{Statistical background subtraction}
\label{SBS}

Even after selecting $UVJ$-passive galaxies on the red sequence, we still expect there to be contamination to a pure cluster sample from foreground and background galaxies. We therefore performed a statistical background subtraction, whereby we measured the background (i.e. number of galaxies per area per mass/mag) in a control field region and subtracted it from the cluster region. In practice, this works by measuring the LF/SMF (section~\ref{KDEM}, equation~\ref{KDEeq}) in both the cluster and control field regions and subtracting the latter from the former. This method is commonly used in the absence of photometric and spectroscopic redshifts, and has proven to be successful in recovering the underlying LF/SMF \citep[e.g.][]{Aragon1993, Andreon2006, Rudnick2009, Mancone2010, Mancone2012, Chan2019, Baxter2021}. Even though we have photometric redshifts and spectra for the massive members, our method is limited by our desire to identify the low-mass and faint passive cluster members. Galaxies belonging to the cluster regions were defined as those within $r < 1$ Mpc from the cluster centre, while galaxies in the control field regions were defined as those at $r > 1$ Mpc from the cluster centre, as shown in Figure~\ref{TCIms}. These definitions were chosen to maximise the number of galaxies in both cluster and control field regions, in order to get accurate measurements of the background without compromising the cluster LFs/SMFs. We discuss how these definitions affect our results at the end of section~\ref{TheLumFuncs}. It is common to have a buffer region in between the cluster and control field regions, but we find this has a negligible effect on our results. 

\subsection{Markov Chain Monte Carlo Schechter fits}
\label{Schfits}

We fit $\hat{\varphi}(\mathcal{M})$ with a Schechter function \citep{Schechter1976} that is characterised by an exponential cut off at the bright or high-mass end and a power law at the faint or low-mass end (with slope $\alpha$), where the transition between the two regimes occurs at the characteristic magnitude/mass $M^{\ast}$. The number density of galaxies at a given magnitude (i.e. luminosity function) is given by

\begin{equation}
\label{Schechter}
\varphi(\mathcal{M}) = 0.4\,\ln(10)\, \varphi^{\ast}\,10^{0.4(M^{\ast} - \mathcal{M})(\alpha + 1)}\,e^{-10^{0.4(M^{\ast} - \mathcal{M})}} ,
\end{equation}
\\
\noindent with the number density of galaxies at a given mass (i.e. SMF) given by

\begin{equation}
\label{SchechterM}
\varphi(\mathcal{M}) = \ln(10)\, \varphi^{\ast}\,10^{(\mathcal{M}-M^{\ast})(\alpha + 1)}\,e^{-10^{(\mathcal{M}-M^{\ast})}} ,
\end{equation}
\\
where $\varphi^{\ast}$ sets the normalisation in both cases. We performed the fitting using the MCMC method so that we can robustly estimate the uncertainties on the Schechter parameters. Since $\varphi^{\ast}$ only sets the normalisation, we ran an initial least-squares fit to determine $\varphi^{\ast}$, and set it as constant throughout the MCMC analysis which determines the best fit for the other parameters $\mathbf{\Theta} = [M^{\ast}, \alpha]$, for which we used weak priors to facilitate the full exploration of parameter space. If we assume that the uncertainties $\hat{\sigma}$ on the measurement $\hat{\varphi}(\mathcal{M})$ are Gaussian, the likelihood that the set of parameters $\mathbf{\Theta}$ produces the observation $\hat{\varphi}(\mathcal{M})$ is given by

\begin{figure*}
    \centering
    \begin{subfigure}{\textwidth}
        \centering
        \includegraphics[width=\textwidth]{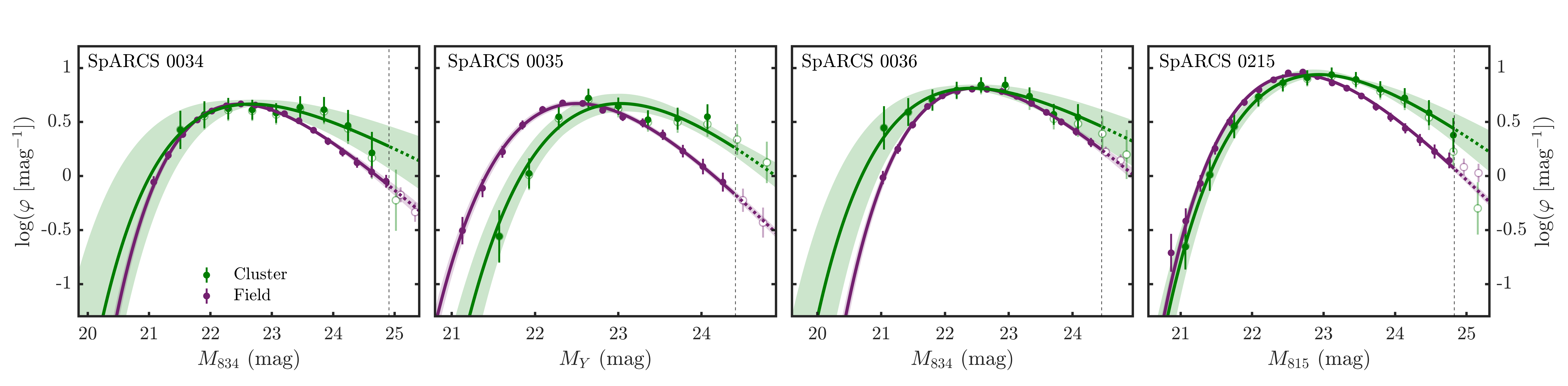}
    \end{subfigure}
    
    \vspace{0.1cm}
    
    \begin{subfigure}{\textwidth}
        \centering
        \includegraphics[width=\textwidth]{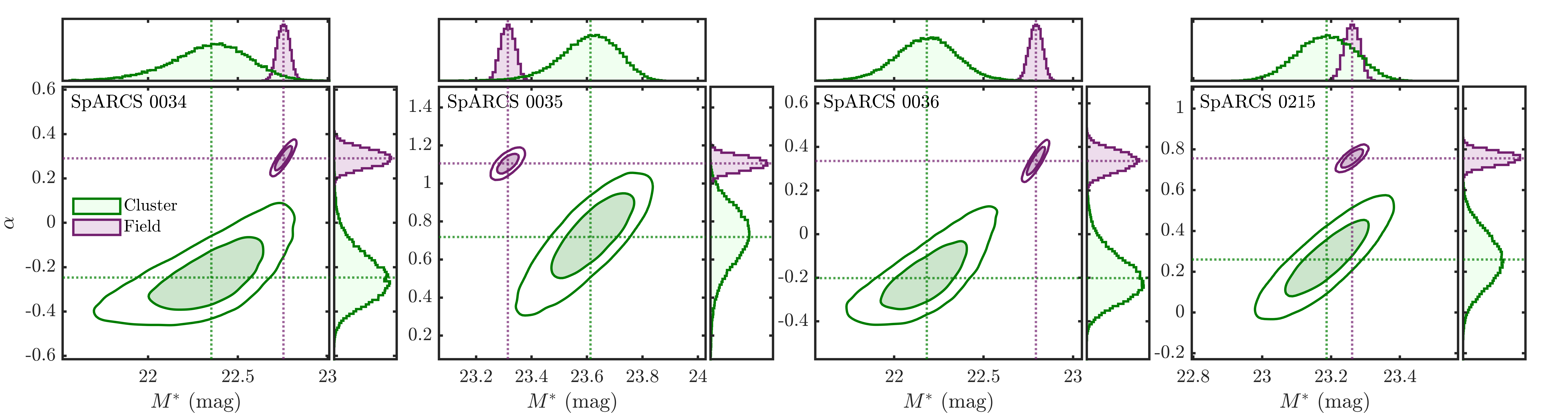}
    \end{subfigure}
    
    \caption{\small{\textit{Top}: The luminosity functions of passive galaxies in the clusters (green) and field (purple). The points are calculated via a Kernel Density Estimation, with the errors deriving from a combination of Monte Carlo simulations and bootstrapping. These points are fit with a Schechter function (equation~\ref{Schechter}) using the MCMC method, where the $M^{\ast}$ and $\alpha$ contours of covariance are shown in the bottom row. The solid lines represent the Schechter fit using the median values of $M^{\ast}$ and $\alpha$, given in Table~\ref{SchParamsMag}, with the shaded region showing the range of 1,000 random samples from within the $1\sigma$ contour of the posterior distribution of the Schechter parameters from the bottom row. In each case, the value of $\varphi^{\ast}$ for the field has been normalised to give the same maximum value as the corresponding cluster. The vertical dashed lines show the $70\%$ completeness limits for the cluster detection images ($F_r$). The Schechter functions were fit to the filled points only. The open green points show the measured LFs of the clusters without correcting for completeness and beyond the completeness limits. The open purple points show the measured field LF beyond completeness limits. \textit{Bottom}: Contours of covariance between the Schechter parameters $M^{\ast}$ and $\alpha$ at the 1 and 2$\sigma$ levels for each cluster (green) and corresponding field (purple). The histograms above and to the side of each contour plot show the number of accepted MCMC steps, with the median values shown by the dotted lines (given in Table~\ref{SchParamsMag}). These median values are used to give the best fits to the LFs in the top row.}}
    \label{LumFuncs}
\end{figure*}

\begin{equation}
\label{Likelihood}
\mathcal{L}(\mathbf{\Theta} \mid \hat{\varphi}(\mathcal{M})) = \prod_{i=1}^{N} \frac{1}{\hat{\sigma}\sqrt{2\pi}} \exp\left[{-\frac{1}{2}\left(\frac{\hat{\varphi}(\mathcal{M}) - \varphi(\mathcal{M},\mathbf{\Theta})}{\hat{\sigma}}\right)^2}\right],
\end{equation}
\\
where $N$ is the number of points in $\hat{\varphi}(\mathcal{M})$, and $\varphi(\mathcal{M},\mathbf{\Theta})$ is the parametric form of the LF/SMF (equation~\ref{Schechter}/\ref{SchechterM}) with parameters $\mathbf{\Theta}$. The log-likelihood is therefore

\begin{equation}
\label{LogLikelihood}
\ln\left[\mathcal{L}\right] = -\frac{1}{2}\sum_{i=1}^{N}\left[\left(\frac{\hat{\varphi}(\mathcal{M}) - \varphi(\mathcal{M},\mathbf{\Theta})}{\hat{\sigma}}\right)^2 +\,\, \ln(2\pi\hat{\sigma}^2)\right].
\end{equation}
\\
We used the Metropolis-Hastings algorithm \citep{Metropolis1953, Hastings1970} to sample from the posterior distribution of the parameters $\mathbf{\Theta}$, given the observed data. This method involves proposing randomly selected new parameter values, and comparing the likelihood values of the proposed and current parameters. We accept the proposed parameters if $\mathcal{R} < \mathcal{L}_{\textrm{proposed}} / \mathcal{L}_{\textrm{current}}$, where $\mathcal{R}$ is a random number drawn from a uniform distribution between 0 and 1. This acceptance ratio is actually calculated using the difference in log-likelihoods (i.e. $\mathcal{R} < \exp[\Delta\ln\mathcal{L}]$, where $\Delta\ln\mathcal{L} = \ln\mathcal{L}_{\textrm{proposed}} - \ln\mathcal{L}_{\textrm{current}}$) to aid stability in the computer's floating point arithmetic. By iterating this process, we build a chain of parameter samples that converge to the target distribution. We began 25 chains at randomly distributed initial points of the parameter space, iterating 5,000 times per chain (but removing the first $10\%$ as burn-in). The best-fit parameter values were calculated as the median of the posterior distribution, with the uncertainties representing the $16-84\%$ confidence intervals.

\section{Results}
\label{Results}

\subsection{The luminosity functions}
\label{TheLumFuncs}

\begin{table}
\centering
\caption{Schechter function (equation~\ref{Schechter}) parameters fit to the luminosity functions of galaxies in the clusters and corresponding field environments. The uncertainties quoted represent the $16-84\%$ confidence interval on each parameter, but do not include systematic uncertainties such as cosmic variance and zeropoint errors.}
\begin{tabular}{wl{0.2\columnwidth} wc{0.165\columnwidth} wc{0.08\columnwidth} wc{0.01\columnwidth} wc{0.165\columnwidth} wc{0.08\columnwidth}}
\hline
 & \multicolumn{2}{c}{Cluster} & & \multicolumn{2}{c}{Field} \\
 \cmidrule(lr){2-3}\cmidrule(lr){5-6}
 Name & $M^{\ast}$ (mag) & $\alpha$ & & $M^{\ast}$ (mag) & $\alpha$ \\
\hline \\ [-2ex]
SpARCS 0034 & 22.35$^{+\,0.24}_{-0.20}$ & -0.25$^{+\,0.10}_{-0.12}$ & & 22.75$^{+\,0.03}_{-0.03}$ & 0.29$^{+\,0.04}_{-0.04}$ \\ [1ex]
SpARCS 0035 & 23.61$^{+\,0.11}_{-0.10}$ & 0.72$^{+\,0.17}_{-0.14}$ & & 23.31$^{+\,0.03}_{-0.03}$ & 1.11$^{+\,0.04}_{-0.04}$ \\ [1ex]
SpARCS 0036 & 22.18$^{+\,0.17}_{-0.16}$ & -0.20$^{+\,0.10}_{-0.12}$ & & 22.79$^{+\,0.04}_{-0.04}$ & 0.34$^{+\,0.05}_{-0.05}$ \\ [1ex]
SpARCS 0215 & 23.19$^{+\,0.08}_{-0.08}$ & 0.26$^{+\,0.13}_{-0.13}$ & & 23.26$^{+\,0.02}_{-0.02}$ & 0.76$^{+\,0.03}_{-0.03}$ \\ [1ex]
\hline
\end{tabular}
\label{SchParamsMag}
\end{table}

The LFs of the four clusters and corresponding field samples are shown in the top row of Figure~\ref{LumFuncs}, with the posterior distributions of the Schechter parameters shown in the bottom row. The median values of the posterior distributions of the Schechter parameters are given in Table~\ref{SchParamsMag}, along with the $16-84\%$ confidence intervals. For three of the clusters, we find a slightly brighter (but similar within uncertainties) characteristic magnitude ($M^{\ast}$) than the corresponding field. This is expected due to the dynamical friction that causes massive halos (and therefore massive, bright galaxies) to fall quickly to the innermost regions of clusters \citep[e.g.][]{Balogh2004, vandenBosch2008, Presotto2012, Roberts2015, Contini2015, Joshi2016, Kim2020}, resulting in an abundance of bright galaxies. Alternatively, \cite{Ahad2024} show that clusters are built from more massive galaxies than the field to begin with. We find the opposite result for SpARCS 0035, where we measure a brighter characteristic magnitude in the field than in the cluster. Compared to the other clusters, we find significantly fewer passive red-sequence galaxies in SpARCS 0035. This means our results for this cluster are less robust than the others.

\begin{figure*}
    \centering
    \begin{subfigure}{\textwidth}
        \centering
        \includegraphics[width=\textwidth]{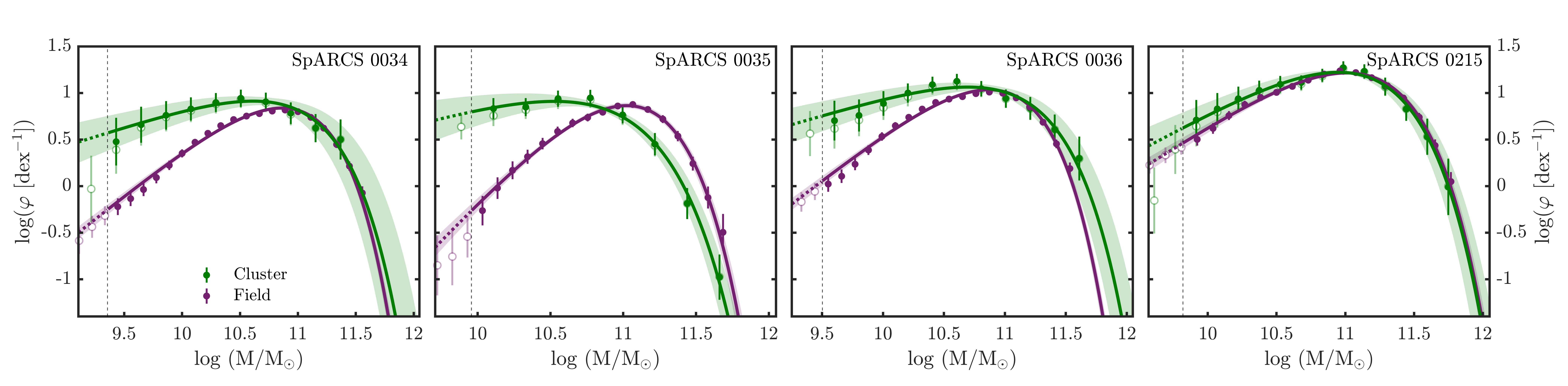}
    \end{subfigure}
    
    \vspace{0.1cm}
    
    \begin{subfigure}{\textwidth}
        \centering
        \includegraphics[width=\textwidth]{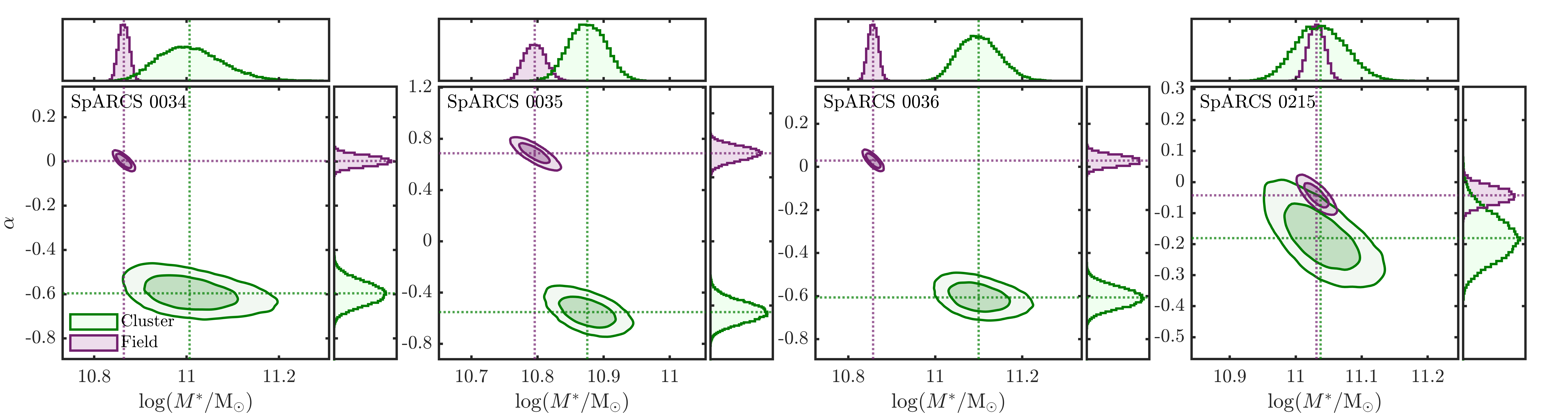}
    \end{subfigure}
    
    \caption{\small{\textit{Top}: The stellar mass functions of passive galaxies in the clusters (green) and field (purple). The points are calculated via a Kernel Density Estimation, with the errors deriving from a combination of Monte Carlo simulations and bootstrapping. These points are fit with a Schechter function (equation~\ref{SchechterM}) using the MCMC method, where the $M^{\ast}$ and $\alpha$ contours of covariance are shown in the bottom row. The solid lines represent the Schechter fit using the median values of $M^{\ast}$ and $\alpha$, given in Table~\ref{SchParamsMass}, with the shaded region showing the range of 1,000 random samples from within the $1\sigma$ contour of the posterior distribution of the Schechter parameters from the bottom row. In each case, the value of $\varphi^{\ast}$ for the field has been normalised match the cluster at the $M^{\ast}$ of the cluster. The vertical dashed lines show the mass limits for the clusters, implicitly derived from the $70\%$ magnitude completeness limit and corresponding mass-luminosity relation (from Figure~\ref{MassMag}). The Schechter functions were fit to the filled points only. The open green points show the measured SMFs of the clusters without correcting for completeness and beyond the completeness limits. The open purple points show the measured field SMF beyond mass limit. \textit{Bottom}: Contours of covariance between the Schechter parameters $M^{\ast}$ and $\alpha$ at the 1 and 2$\sigma$ levels for each cluster (green) and corresponding field (purple). The histograms above and to the side of each contour plot show the number of accepted MCMC steps, with the median values shown by the dotted lines (given in Table~\ref{SchParamsMass}). These median values are used to give the best fits to the SMFs in the top row.}}
    \label{MassFuncs}
\end{figure*}

We find that all clusters exhibit a gradual decrease of passive red-sequence galaxies towards the faint end, with all clusters showing $\alpha > -0.25$. This decrease, however, is not as dramatic as the decrease in the corresponding field samples, which all show $\alpha > 0.29$. These differences in the faint-end slope are significant, with two-tailed $p$-values less than 0.02 across all four clusters. Ultimately, this means that we find relatively more faint passive galaxies in these clusters than in the field. While these results also seem to suggest some redshift evolution in the faint-end slope, where the steepness increases with redshift, it would not be valid to directly compare the Schechter parameters between clusters. This is due to the inhomogeneity in the selection of galaxies between clusters, as well as the LF being measured using different filters. The only fair comparison is between a cluster and its corresponding field sample, where we do find significant differences.

\cite{Werner2022} show that the population of galaxies in the outskirts of clusters is not the same as a true field population, where they find an enhancement in the number of infalling massive quenched galaxies. In addition, pre-processing means that the cluster surroundings are likely to have a higher quenched fraction than a true field sample. To see whether our results are significantly affected by this, we varied the cluster/control field boundary used for the background subtraction between $0.5 < r < 1.5$ Mpc. This upper limit is determined by the extent of our data, meaning it is unlikely we ever reach a true field sample. We find our main conclusions remain unchanged, but for SpARCS 0035, the Schechter fits are sensitive to this definition. This is due to the low number of passive red-sequence galaxies in this cluster, which results in the LF and SMF being sensitive to the inclusion or exclusion of certain galaxies. While results for this cluster must be taken with caution, we still consistently measure a steeper slope in the LFs and SMFs at the faint and low mass ends in the field compared to the cluster, even if the exact value of $\alpha$ varies significantly.

\subsection{The stellar mass functions}
\subsubsection{Individual SMFs}
\label{IndSMFS}

As SMFs offer deeper insights into the mass assembly and evolutionary processes of galaxies than LFs, we also show the SMFs of the four clusters and corresponding field samples are shown in the top row of Figure~\ref{MassFuncs}, with the posterior distributions of the Schechter parameters shown in the bottom row. The median values of the posterior distributions of the Schechter parameters are given in Table~\ref{SchParamsMass}, along with the $16-84\%$ confidence intervals. For three of the clusters, we find a slightly larger characteristic mass ($M^{\ast}$; though similar within uncertainties) than the corresponding field. This is expected due to dynamical friction or possibly through the cluster's assembly from massive galaxies in the first place (as explained in section~\ref{TheLumFuncs}),  both of which result in an abundance of massive galaxies. While we find a slight difference at the massive end, systematic uncertainties in the measurement of $M^{\ast}$ of $\sim0.3$ dex \citep[see e.g.][]{Marchesini2009}  would put our results in agreement with \cite{vanderBurg2013, vanderBurg2020}. Similar to the LFs, we find the opposite result for SpARCS 0035, where we measure a higher characteristic mass in the field than in the cluster. As explained above, our results for this cluster are less robust than the others. Also important to note is that due to the differing mass limits between clusters, comparisons between the Schechter parameters of the individual clusters should be taken with caution.

\begin{figure*}
    \centering
    {\includegraphics[width=0.55\textwidth]{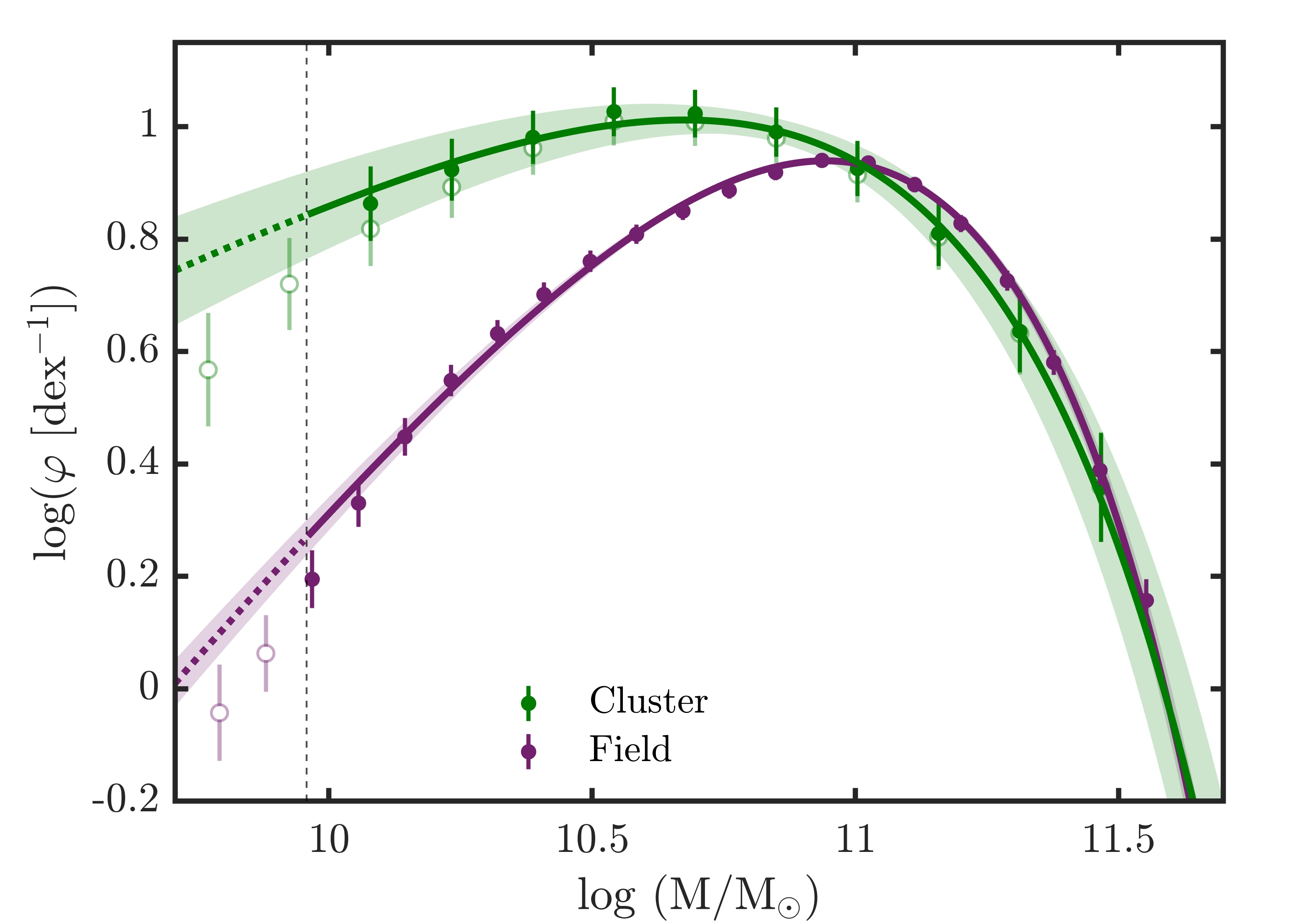}}
    {\includegraphics[width=0.4\textwidth]{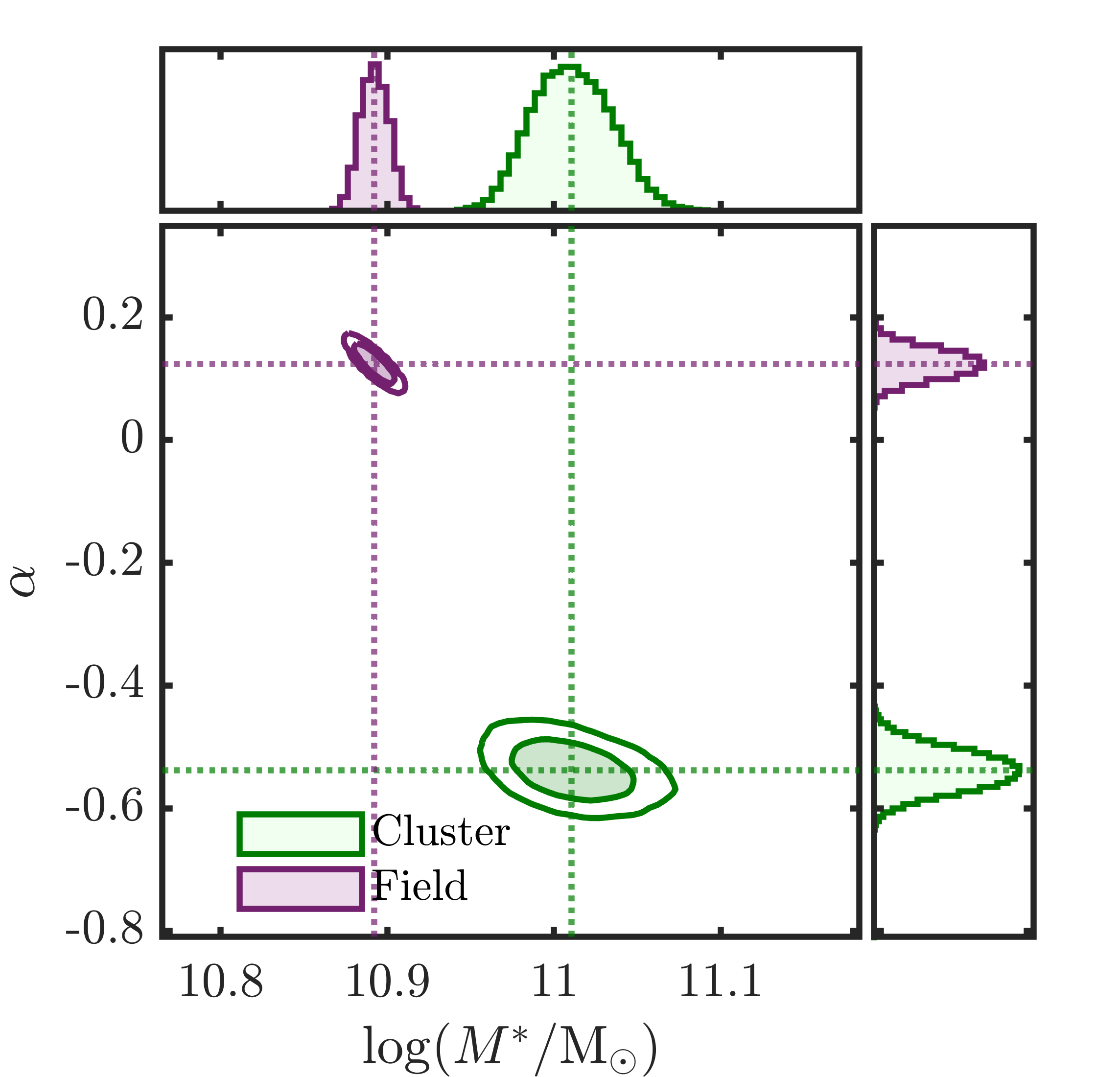}}  
    \caption{\small{\textit{Left}: The composite stellar mass function of passive galaxies in the clusters (green) and field (purple). The points are calculated by combining the individual SMFs from Figure~\ref{MassFuncs} and propagating their errors. These points are fit with a Schechter function (equation~\ref{SchechterM}) using the MCMC method, where the $M^{\ast}$ and $\alpha$ contours of covariance are shown in right panel. The solid lines represent the Schechter fit using the median values of $M^{\ast}$ and $\alpha$, given in Table~\ref{SchParamsMass}, with the shaded region showing the range of 1000 random samples from within the $1\sigma$ contour of the posterior distribution of the Schechter parameters from the right panel. The value of $\varphi^{\ast}$ for the field has been normalised match the cluster at the $M^{\ast}$ of the cluster. The vertical dashed line shows the maximum mass limit of the four clusters. The Schechter functions were fit to the filled points only. The open green points show the measured SMFs of the clusters without correcting for completeness and beyond the completeness limit. The open purple points show the measured field SMF beyond the mass limit. \textit{Right}: Contours of covariance between the Schechter parameters $M^{\ast}$ and $\alpha$ at the 1 and 2$\sigma$ levels for the clusters (green) and field (purple). The histograms above and to the side of the contour plot show the number of accepted MCMC steps, with the median values shown by the dotted lines (given in Table~\ref{SchParamsMass}). These median values are used to give the best fits to the SMFs in the left panel.}}
    \label{CompSMF}
\end{figure*}

We find that all clusters exhibit a gradual decrease of passive red-sequence galaxies towards the low-mass end, with all clusters showing $\alpha > -0.7$. This decrease, however, is not as dramatic as the decrease in the corresponding field samples, which all show $\alpha > -0.1$. For SpARCS 0034, 0035 and 0036, these differences in the low-mass-end slope are highly significant, with two-tailed $p$-values less than $1\times10^{-28}$. For SpARCS 0215, the difference is insignificant, with a two-tailed $p$-value of 0.07. While this is insignificant, we do find a significant difference in the faint end slope of the corresponding LF for this cluster. The $F_r$ bands used to construct the LFs roughly probe rest-frame V-band light, which is not a reliable mass tracer for galaxies that have only recently shut down their star formation. It is therefore not necessary that both LF and SMF agree. However, while there is a discrepancy in the significance of the difference between cluster and field between LF and SMF, it is a fairly small discrepancy (the $p$-value for the SMF borders on the threshold of statistical significance). Our construction of the composite SMF in section~\ref{compSMFs} aims to alleviate these issues.

We also note here the more significant differences between the SpARCS 0034, 0035 and 0036 clusters and corresponding field $\alpha$ values for the SMFs compared to the LFs. The SMFs are based on $12+$ band fitting, as opposed to just one measurement for the LFs, which may increase the precision of the mass measurements. However, there are other systematics introduced \citep[see][for a detailed assessment]{Marchesini2009}. We therefore do not believe this increase is caused by any physical process, but rather is as a result of the assumptions and uncertainties that go into the mass estimates.

\begin{table}
\centering
\caption{Schechter function (equation~\ref{SchechterM}) parameters fit to the SMFs of galaxies in the clusters and corresponding field environments. The uncertainties quoted represent the $16-84\%$ confidence interval on each parameter, but do not include systematic uncertainties such as cosmic variance and zeropoint errors.}
\begin{tabular}{wl{0.2\columnwidth} wc{0.165\columnwidth} wc{0.08\columnwidth} wc{0.01\columnwidth} wc{0.165\columnwidth} wc{0.08\columnwidth}}
\hline
 & \multicolumn{2}{c}{Cluster} & & \multicolumn{2}{c}{Field} \\
 \cmidrule(lr){2-3}\cmidrule(lr){5-6}
 Name & $\textrm{log}(M^{\ast}/$M$_{\odot})$ & $\alpha$ & & $\textrm{log}(M^{\ast}/$M$_{\odot})$ & $\alpha$ \\
\hline \\ [-2ex]
SpARCS 0034 & 11.01$^{+\,0.05}_{-0.05}$ & -0.60$^{+\,0.05}_{-0.05}$ & & 10.86$^{+\,0.01}_{-0.01}$ & 0.00$^{+\,0.02}_{-0.02}$ \\ [1ex]
SpARCS 0035 & 10.88$^{+\,0.03}_{-0.03}$ & -0.55$^{+\,0.08}_{-0.08}$ & & 10.80$^{+\,0.02}_{-0.02}$ & 0.69$^{+\,0.06}_{-0.05}$ \\ [1ex]
SpARCS 0036 & 11.10$^{+\,0.05}_{-0.05}$ & -0.61$^{+\,0.04}_{-0.05}$ & & 10.86$^{+\,0.01}_{-0.01}$ & 0.04$^{+\,0.02}_{-0.02}$ \\ [1ex]
SpARCS 0215 & 11.04$^{+\,0.04}_{-0.04}$ & -0.18$^{+\,0.07}_{-0.07}$ & & 11.03$^{+\,0.01}_{-0.01}$ & -0.04$^{+\,0.03}_{-0.03}$ \\ [1ex]
\hline \\ [-2ex]
Composite & 11.01$^{+\,0.02}_{-0.03}$ & -0.54$^{+\,0.03}_{-0.03}$ & & 10.89$^{+\,0.02}_{-0.02}$ & 0.12$^{+\,0.02}_{-0.02}$ \\ [1ex]
\hline
\end{tabular}
\label{SchParamsMass}
\end{table}

\subsubsection{Composite SMF}
\label{compSMFs}

In addition to the individual SMFs described in section~\ref{IndSMFS}, we also produce a composite SMF. Combining multiple clusters into one LF or SMF is commonly done to increase the total number of galaxies, and therefore improve statistics \citep[e.g.][]{Vulcani2011, vanderBurg2013, vanderBurg2020, Chan2019}. This allows us to constrain the cluster-average Schechter parameters $M^{\ast}$ and $\alpha$ to a higher accuracy. As the magnitudes of galaxies in the clusters are measured in different filters, it is not feasible to produce a composite  LF. It is possible to create a composite SMF, however. 

The composite SMF was created by measuring the individual SMFs as before (i.e. sections~\ref{KDEM},~\ref{SBS} and~\ref{Schfits}), except this time evaluating the KDE at the same points for each of the clusters, allowing a mean to be calculated, and also using the most conservative mass limit of the four clusters: log(M/M${_\odot}$) $ = 9.96$. It was done this way so that the background subtraction and completeness correction for each cluster could be performed. While this may seem to not reach much deeper than the log(M/M${_\odot}$) $ = 10$ depth of \cite{vanderBurg2013}, we in fact have many more faint/low-mass galaxies making up the SMF (as shown by Figure~\ref{CMDs}). This is because \cite{vanderBurg2013} perform a membership correction, which artificially lowers their mass limit, whereas we actually detect these lower mass galaxies. The spacing of the points was chosen to not artificially reduce uncertainties of the MCMC fits by using the \cite{Silverman1986} rule-of-thumb bandwidth for the combined sample (as opposed to the individual samples from before). The errors on the individual SMFs are propagated into the composite SMF. This method is performed for both the cluster and field samples, where each field sample is treated as separate and then combined in the same fashion described above. 

The composite SMF for the clusters and field is shown in Figure~\ref{CompSMF}, along with the posterior distribution of the Schechter parameters. The median values of the posterior distribution of the Schechter parameters are shown in Table~\ref{SchParamsMass}, along with the $16-84\%$ confidence intervals. Similarly to the individual SMFs, we find the clusters have a slightly larger characteristic mass compared to the field, which have $\textrm{log}(M^{\ast}/$M$_{\odot}) = 11.01^{+\,0.02}_{-0.03}$ and $\textrm{log}(M^{\ast}/$M$_{\odot}) = 10.89^{+\,0.02}_{-0.02}$, respectively. Although, considering cosmic variance and zero point uncertainties (which are not included in the statical uncertainties quoted) these are in approximate agreement. The clusters also exhibit a gradual decrease of passive red-sequence galaxies towards the low-mass end, with $\alpha = -0.54^{+\,0.03}_{-0.03}$. The field, however, has a much sharper decrease of passive red-sequence galaxies towards the low-mass end, with $\alpha = 0.12^{+\,0.02}_{-0.02}$. The significance of this difference in the low-mass end slope between the clusters and field is extremely high, with a two-tailed $p$-value less than $1\times10^{-74}$.

\section{Discussion}
\label{Disc}

\subsection{Literature comparisons}
As an extension of the work presented in \cite{vanderBurg2013}, we first compare our results to theirs. In their composite cluster SMF of passive galaxies, they measure a characteristic mass of log$(M^{\ast}/$M$_{\odot}) = 10.71^{+\,0.04}_{-0.10}$, compared to ours of log$(M^{\ast}/$M$_{\odot}) = 11.01^{+\,0.02}_{-0.03}$. While our results are not too dissimilar, the difference in the characteristic masses is statistically significant. Cosmic variance of massive galaxies may be the main reason for this discrepancy since our composite SMF consists of only 4 of the 10 clusters that are used in \cite{vanderBurg2013}.

Our measurement of the characteristic mass is consistent with several other works, such as \cite{Tomczak2017} which is based on the ORELSE survey \citep{Lubin2009} and studies different density regions at $0.55 < z < 1.3$. In their highest two density bins, they measure log$(M^{\ast}/$M$_{\odot}) = 11.07\pm0.13$ and log$(M^{\ast}/$M$_{\odot}) = 11.04\pm0.09$, for passive galaxies. The results from \cite{Davidzon2016} are also consistent with ours. They measure log$(M^{\ast}/$M$_{\odot}) = 10.97^{+\,0.09}_{-0.07}$ for passive galaxies in their high density regions, though this is measured at lower redshifts of $0.65 < z < 0.80$.

\begin{figure}
    \centering    {\includegraphics[width=\columnwidth]{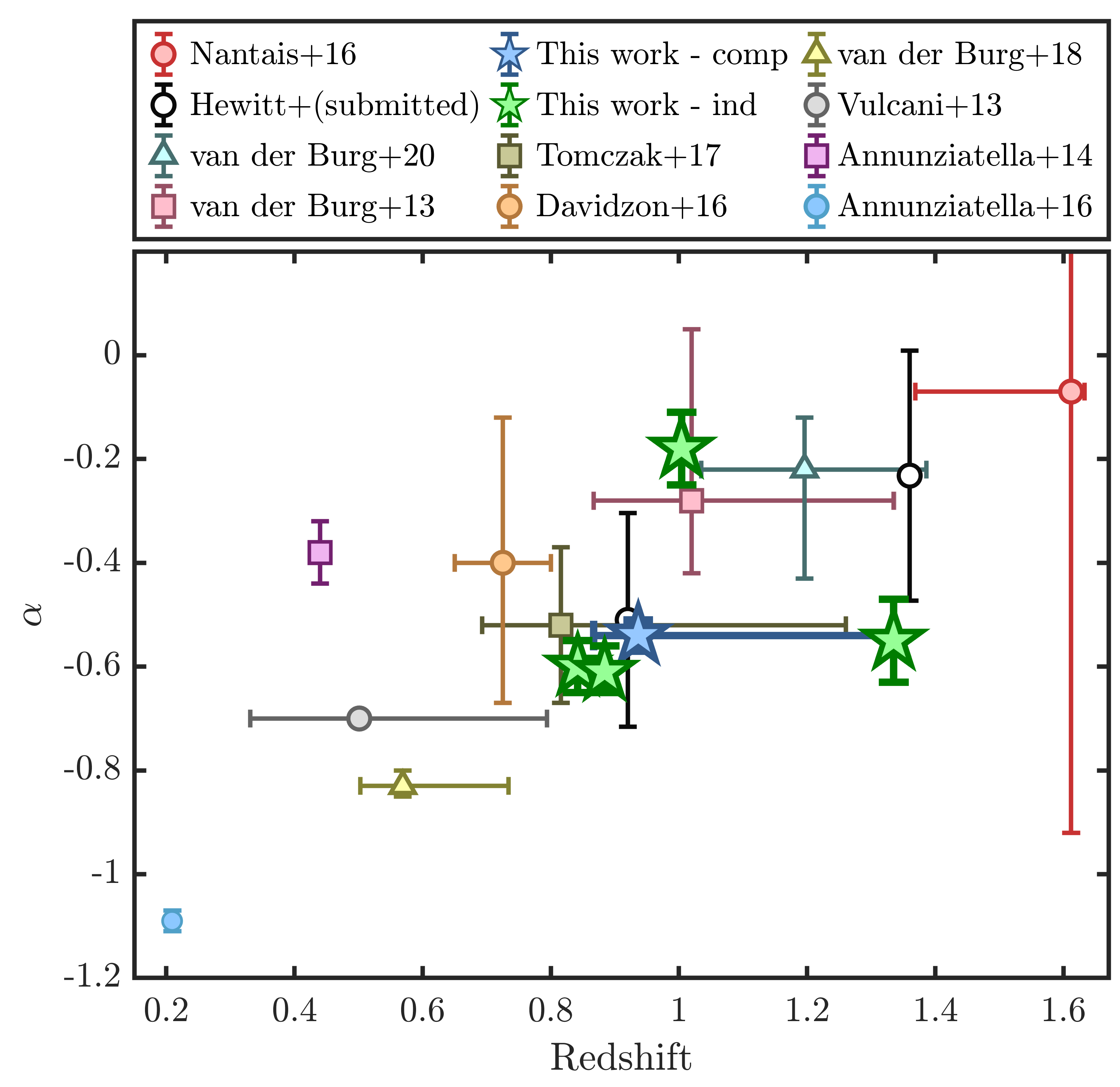}} 
    \caption{\small{The gradient of the low-mass end of the SMF ($\alpha$) of passive cluster galaxies as a function of redshift, for different studies. The works we compare to in redshift order are: \protect\cite{Annunziatella2016}, \protect\cite{Annunziatella2014}, \protect\cite{Vulcani2013}, \protect\cite{vanderBurg2018}, \protect\cite{Davidzon2016}, \protect\cite{Tomczak2017}, \protect\cite{vanderBurg2020}, \protect\cite{vanderBurg2013}, Hewitt et al. (submitted), and \protect\cite{Nantais2016}. The green stars show our new measurements for the individual clusters, with the blue star showing our measurement for the composite SMF. The redshifts for the SpARCS~0034 and SpARCS~0036 clusters studied in this work have been shifted slightly in this figure for visualisation purposes. For the studies based on multiple clusters, the horizontal error bars show the entire redshift range of their clusters, with the point showing the median redshift of the clusters studied. Although a robust comparison between the different results is difficult due to the heterogeneous galaxy selection methods, there is a broad trend towards higher alpha (i.e. flatter slopes) at higher redshift.}}
    \label{redalph}
\end{figure}

Focusing on the slope of the low-mass end of the SMF for passive galaxies, \cite{vanderBurg2013} measure $\alpha = -0.28^{+\,0.33}_{-0.14}$, compared to ours of $\alpha = -0.54^{+\,0.03}_{-0.03}$. Again, these results differ significantly and there are a number of reasons why this may be. First, our methods differ since we select galaxies on the red sequence which \cite{vanderBurg2013} do not. On top of this, we use different methods to measure the SMFs themselves. Finally, our goal was to extend the work of \cite{vanderBurg2013} to lower masses, which on average we do by 0.67 dex per cluster (though, our completeness limits are not directly comparable). Given this, and the other reasons above, it is not surprising that the low-mass slopes we measure in the clusters differ slightly to that of \cite{vanderBurg2013}. On the other hand, using largely the same cluster sample as \cite{vanderBurg2020} but an alternative method, Hewitt et al. (submitted) measure the SMF of 17 $1 < z < 1.5$ clusters. By fitting unbinned data in a Bayesian approach, they account for the varying mass limits of each cluster without the upweighting applied by \cite{vanderBurg2020}. Also, by explicitly accounting for variation in SMF with clustercentric radius, their conclusion is that the \cite{vanderBurg2020} approach somewhat overestimates the precision of their SMFs at the lowest masses. With this more sophisticated approach, they are finding results consistent with ours (as shown in Figure~\ref{redalph}).

Our low-mass slope is also fairly consistent with other works in the literature such as \cite{Tomczak2017} and \cite{Davidzon2016}, as can be seen in Figure~\ref{redalph}  where we compare our measured $\alpha$ values for the passive cluster SMFs to others from the literature. In their highest two density bins,  \cite{Tomczak2017} measure $\alpha = -0.63\pm0.2$ and $\alpha = -0.52\pm0.15$, while \cite{Davidzon2016} measure $\alpha = -0.40^{+\,0.28}_{-0.27}$ in their high density regions, though this is measured at lower redshifts of $0.65 < z < 0.80$. Hence we conclude that our results largely agree with previous results in the literature. Another recent work that is consistent with ours is \cite{Hamadouche2024}, who find strong evidence of environmental quenching of low-mass galaxies out to $z\sim2$, though do not specifically look at clusters and also fit double Schechter functions making direct comparisons tricky.

\begin{figure*}
    \centering    {\includegraphics[width=0.9\textwidth]{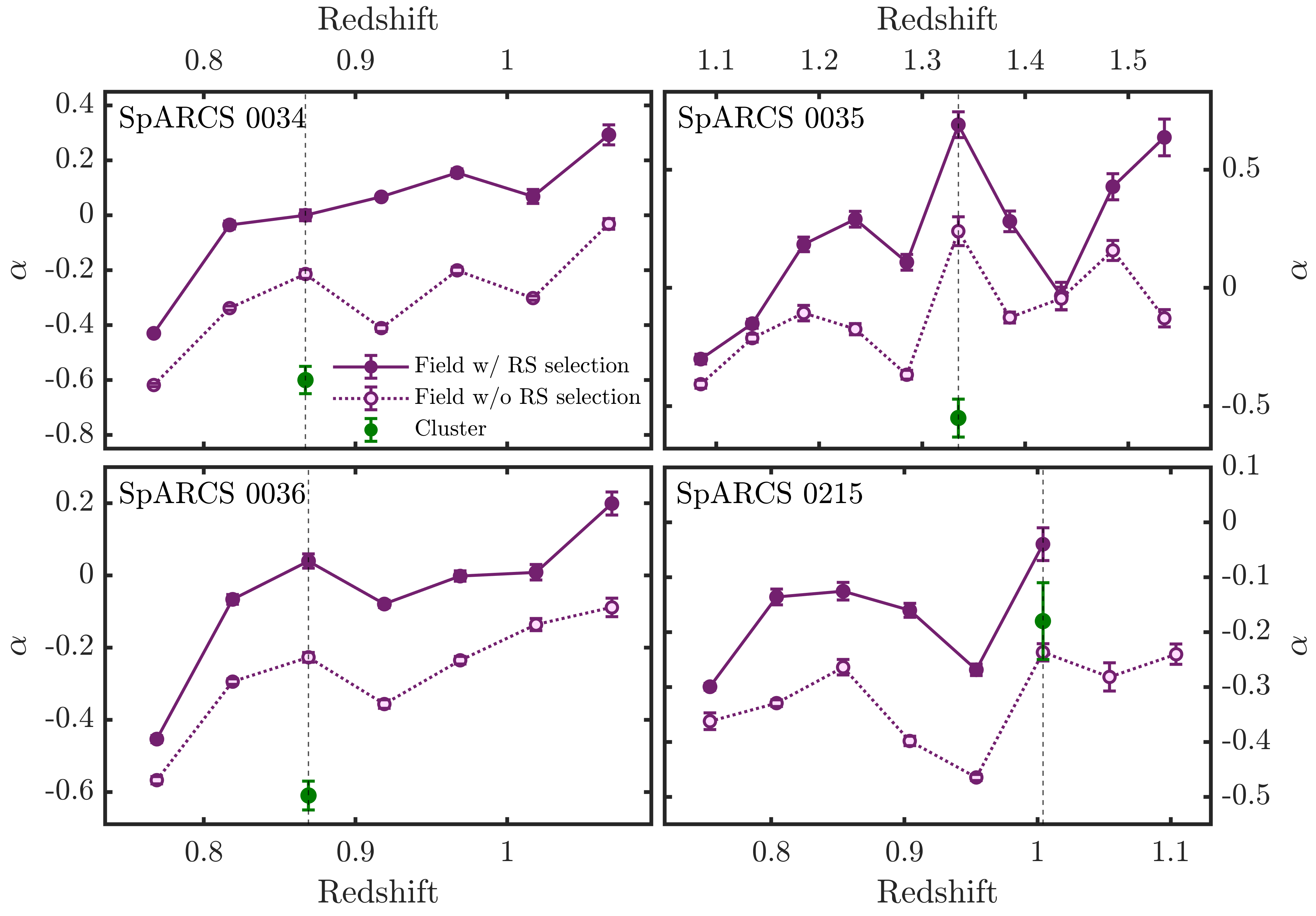}} 
    \caption{\small{The gradient of the low-mass end of the SMF ($\alpha$) of passive galaxies as a function of redshift. The Schechter parameter $\alpha$ is measured using the MCMC method with error bars representing 1$\sigma$. The solid purple line represents the $\alpha$ values for field galaxies selected on the red-sequence in non overlapping redshift bins of width $\Delta z = 0.05$. The dashed purple line represents the same as the solid line except for the red-sequence selection. In both cases, the passive galaxies are selected using $UVJ$ colours. The green point represents the $\alpha$ value measured for the corresponding cluster, with the vertical dashed line showing the redshift of the cluster.}}
    \label{FieldAlphas}
\end{figure*}


The largest difference between our results and the results of \cite{vanderBurg2013} and \citet{vanderBurg2020} are the low-mass end slopes of the passive field SMFs, where we measure $\alpha = 0.12^{+\,0.02}_{-0.02}$ compared to theirs of $\alpha = -0.43^{+\,0.02}_{-0.04}$ and $\alpha = -0.22^{+\,0.09}_{-0.09}$, respectively. The field samples used in the aforementioned works are based on the wide-field NIR survey UltraVISTA \citep{McCracken2012} in the COSMOS field, with the first selecting galaxies in $0.85 < z < 1.2$ and the second in $1.0 < z < 1.4$. We take a slightly different approach and select galaxies in much narrower redshift slices ($\Delta z = 0.05$) around the cluster redshifts. As narrower redshift slices are more sensitive to cosmic variance, we measured the low-mass end slopes in the field as a function of redshift. This redshift range was limited to ensure the 4000\,{\AA} break of galaxies falls within $F_r$ and $F_b$. This was done for each of the VLT filter combinations and red-sequence selections that corresponds to each cluster. The solid lines in Figure~\ref{FieldAlphas} show a clear redshift evolution of the low-mass end slopes of passive field galaxies. This is consistent with the work of \cite{Weaver2023}, and shows a steady, relative increase of low-mass passive galaxies over time. Importantly, we find our choice of redshift slice for the field samples does not change our conclusions, as the relative difference between the low-mass end slope for the field and clusters remains qualitatively consistent throughout. 

While Figure~\ref{FieldAlphas} shows the results for the field samples are not skewed by cosmic variance, we do find systematically higher $\alpha$ values than previous works \citep[e.g.][]{Muzzin2013, vanderBurg2013, vanderBurg2020, Weaver2023}. Therefore, in Figure~\ref{FieldAlphas}, we also show the redshift evolution of the low-mass end slope for the field samples when we do not make the red-sequence selection. This brings the method more in line with previous works which also do not make a red-sequence selection. We find a mean offset between the low-mass end slope when selecting galaxies on the red-sequence and not of 0.27. This offset shows that the higher $\alpha$ values we measure in the field compared to other works is mostly caused by the red-sequence selection. A smaller contribution to this difference may be caused by the larger redshift slices used in other works, which can push the $\alpha$ value down due to the inclusion of many lower redshift galaxies (which, as shown in Figure~\ref{FieldAlphas}, have a much lower $\alpha$). We can also see here why we measure a much higher $\alpha$ for the field sample of SpARCS 0035 compared to the other cluster's field samples. There happens to be a spike in $\alpha$ at the redshift of the cluster which deviates from the general trend in redshift. This is likely caused by cosmic variance, where there happens to be structure (or lack of) in this redshift slice. Assuming this is cosmic variance, an $\alpha$ value on the general trend with redshift would be much more consistent with the other cluster's field samples.

In this work, the field environment reflects a representative or average part of the Universe which, accordingly, includes both underdensities and overdensities of galaxies alike. Previous studies such as \cite{Kawinwanichakij2017} and \cite{Papovich2018}, however, define their field environments as the lowest density quartiles, attempting to maximise the chance of seeing an environmental difference \citep[see also][]{Cooper2010}. According to the \cite{Peng2010} model, this low-density environment would have a scarcity of low-mass passive galaxies, with only internal mechanisms acting to quench galaxies. Including overdensities in the field, we expect environmental quenching to at least somewhat contribute to the total passive population, and possibly even be the sole contributor to the low-mass passive population\footnote{This is seen in simulations where even at $z\sim2$, there are quenched cluster satellites with mass $10^{9}$ - $10^{10}$ M$_{\odot}$ but no quenched field centrals of the same mass \citep{Ahad2024}.}. Correcting for this inclusion of environmental quenching in the field would only lead to a starker difference in the faint and low mass end slopes of the LFs and SMFs between cluster and field.

Another notable point regarding the field is that we have used data obtained from observations different from those of our cluster sample. This is not ideal when performing a direct comparison between the two environments. We have, however, taken numerous steps to bring the field data in line with the cluster data, and believe the comparison is fair and valid. Improvements in future studies can be to make as much of an attempt to use homogeneous cluster and field data as possible.

\subsection{Environmental quenching at $z\sim1$}

Unlike our results, \cite{vanderBurg2013, vanderBurg2020} find almost indistinguishable shapes of the SMFs of passive galaxies between clusters and the field at $z\sim1$ (down to their mass limits of $10^{10}$ M$_{\odot}$ and $10^{9.7}$ M$_{\odot}$, for the 2013 and 2020 works, respectively). The authors argue that their results show that quenching mechanisms at $z\sim1$ work differently from what is observed in the local Universe, and that the environmental excess quenching at the higher redshifts is strongly dependent on stellar mass. \cite{Webb2020} show that galaxies within $z\sim1$ clusters have slightly earlier formation times compared to the field, which \cite{vanderBurg2020} postulate could explain the mass-dependent quenched fraction excess they measure. With our seemingly contradictory results, we discuss the applicability of traditional quenching models at higher redshifts, and whether, like \cite{vanderBurg2013, vanderBurg2020}, our results require an explanation different to that of the quenching in the local Universe.

\begin{figure}
    \centering    {\includegraphics[width=\columnwidth]{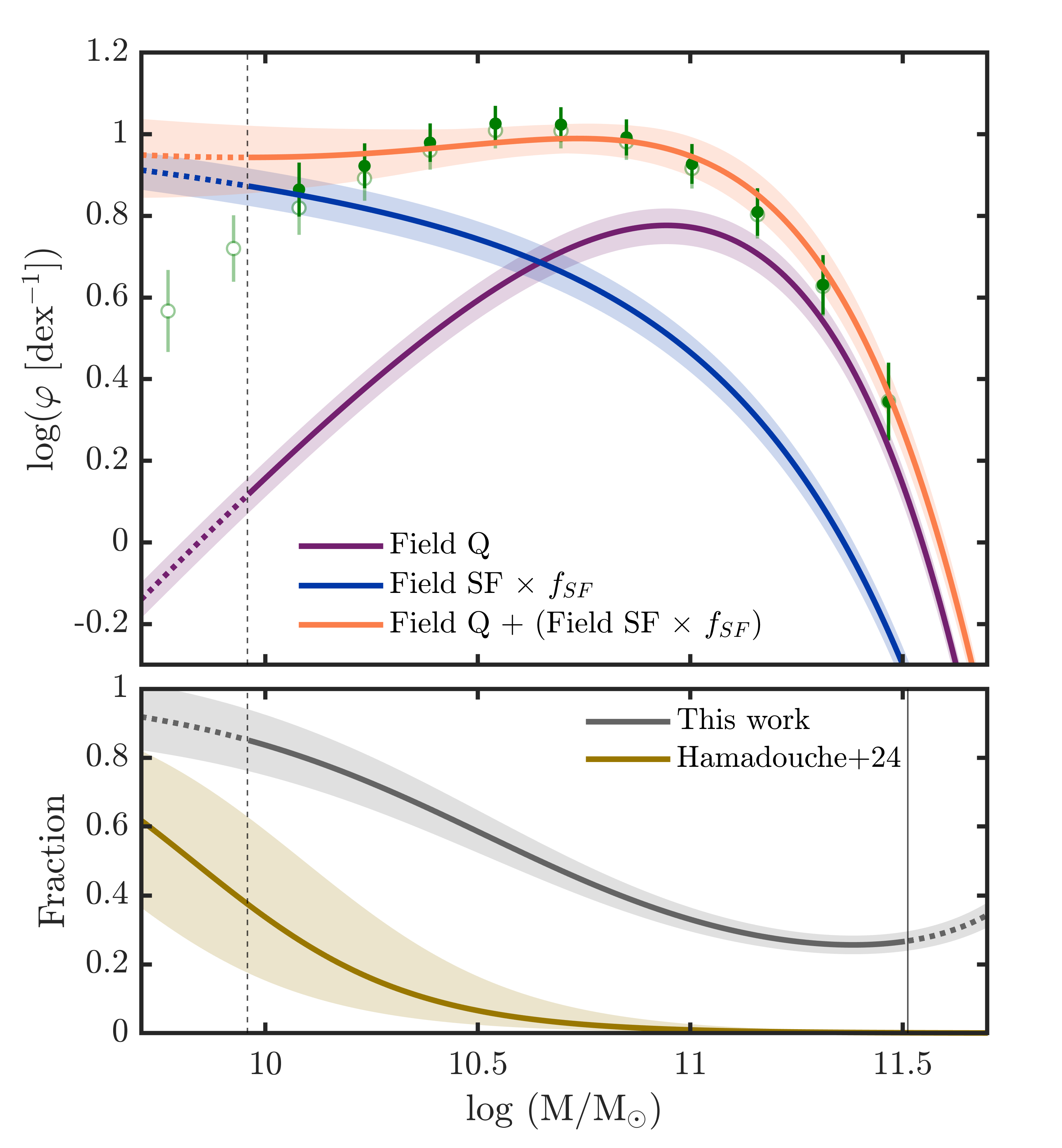}} 
    \caption{\small{\textit{Top}: The double Schechter function (orange) to the composite SMF of passive galaxies in the clusters (green points). The double Schechter function is comprised of the single Schechter functions fit to passive and star-forming field galaxies, whose relative contributions are shown by the purple and blue curves respectively, and are fit using the MCMC method. The green points are the same as in Figure~\ref{CompSMF}. The purple curve has the same shape as in Figure~\ref{CompSMF}, but is renormalised by its contribution to the double Schechter function. The solid curves, shaded regions and vertical dashed line all have the same meaning as in Figure~\ref{CompSMF}. We measure that the fraction of star-forming field galaxies that would need to be quenched to match the passive cluster SMF as $f_{SF} = 25\pm5\%$. \textit{Bottom}: The fractional contribution of the star-forming field population to the double Schechter function from the panel above, as a function of mass (grey). Taking our assumptions that the star-forming field population represent the galaxies that would be environmentally quenched in a cluster, this shows what fraction of passive galaxies are quenched environmentally, as opposed to internally. We also show the equivalent fraction for the double Schechter function that is fit to the passive galaxy population in \protect\cite{Hamadouche2024}. The solid vertical line represents the limit at the high-mass end for which we perform the fitting. This value is the minimum of the highest mass galaxy across the four clusters.}}
    \label{CompSMFSF}
\end{figure}

If we assume that the separability of mass and environmental quenching holds in $z\sim1$ clusters, as per the \cite{Peng2010} model, we would expect the passive cluster SMF to be comprised of the equivalent mass-quenched population measured in the field, with an additional independent environmentally quenched population. This environmentally quenched population in the clusters would arise from the quenching of the star-forming population in the field. As this star forming population has an abundance of low-mass galaxies (see e.g. the top panel of Figure~\ref{CompSMFSF}), its quenching would dramatically alter the low-mass end of the passive SMFs, assuming the quenching is independent of stellar mass. This would manifest as a relative upturn at the low-mass end of the SMF in the clusters compared to the field. 

We therefore fit the composite passive cluster SMF with a combination of the mass-quenched passive field galaxies and would-be quenched star forming field galaxies through the addition of two Schechter functions. The Schechter parameters $M^{\ast}$ and $\alpha$ are measured for both the passive and star forming populations in the field (see fits in Figure~\ref{CompSMFSF}), where the star forming galaxy sample is created using the same selection as the passive sample, except with the opposite $UVJ$ criteria (equations~\ref{uvjcrit}) and no red-sequence selection. The relative contributions of each Schechter function is determined by their respective $\varphi^{\ast}$ value. It is these normalisation parameters that are fit to the cluster SMF via an MCMC method.  

The top panel of figure~\ref{CompSMFSF} shows this fit along with the SMFs of the passive and star-forming populations in the field, normalised by their relative contributions. We show that the passive cluster SMF can be modeled with a double Schechter function which combines the passive and star-forming populations measured in the field. If we assume the entirety of the passive population in the field contributes to the passive cluster population, the fraction of the star-forming field population in the same volume that needs to be quenched to match the shape of the cluster SMF is measured to be $f_{SF} = 25\pm5\%$. The bottom panel of figure~\ref{CompSMFSF} shows the fraction of galaxies that are quenched environmentally in this scenario. At higher masses, we find that internal quenching dominates, and at lower masses, environmental quenching dominates. At log(M/M${_\odot}$) $ = 11.5$, $\sim70\%$ of passive galaxies are internally quenched and only $\sim30\%$ environmentally quenched, but at log(M/M${_\odot}$) $ = 10$, $\sim80\%$ are environmentally quenched. The enhancement of passive galaxies at the low-mass ends of the cluster SMFs compared to the field shows the impact of environmental quenching. Therefore, quenching processes that act in the clusters either do not happen in low-density regions, or are enhanced in clusters. 

\cite{Hamadouche2024} measure SMFs for star-forming and quiescent galaxies without differentiating by environment, using the JWST PRIMER survey. The double Schechter function they fit to their passive population will be comprised of the internally quenched galaxies in the field, and environmentally quenched galaxies in high-density regions - though dominated by the more numerous massive field galaxies. Assuming separability, we can directly compare our results. In the first part of the double Schechter function they fit to the passive population at $0.75 < z < 1.25$, which we assume comprises the internally quenched field galaxies, they measure $\alpha = 0.19\pm0.45$. In the second part, which we assume comprises the environmentally quenched galaxies, they measure $\alpha = -1.53\pm0.2$. For the direct comparison of the shapes, we use the passive field population to compare with the first part of their fit, and the star-forming (would-be quenched) field population to compare with the second part of their fit. For these two populations, we measure $\alpha = 0.12^{+\,0.02}_{-0.02}$ and $\alpha = -1.10^{+\,0.01}_{-0.01}$, respectively. The internally quenched populations therefore match well, though the low-mass end slopes of the environmentally quenched populations differ by $\sim2\sigma$. 

Next, we compare the relative contributions of the two populations. In the bottom panel of Figure~\ref{CompSMFSF}, we show the fractional contribution of one part of the double Schechter function to the total, for both our work and that of \cite{Hamadouche2024}. Here we assume this fraction represents the fraction of galaxies that are quenched environmentally as opposed to internally. We find a similar trend to \cite{Hamadouche2024}, in which environmental quenching dominates towards the lower masses while internal quenching dominates at the high-mass end. However, we find a much larger contribution from the environmentally quenched galaxies overall. It is not surprising that our measurement is higher, as it is based specifically on clusters, as opposed to the entire quenched population, meaning we force the contribution from the environmentally quenched population to be higher. Overall, our results are in strong agreement with \cite{Hamadouche2024}, and largely support traditional quenching models.

While our results fit better with traditional quenching models compared to the works of \cite{vanderBurg2013, vanderBurg2020}, we only study a very small sample of clusters. Much larger samples of clusters are needed to truly understand the nature of high-redshift environmental quenching. Our findings, whilst statistically robust, may not represent the entire $z\sim1$ cluster population. In addition to this, deeper studies are needed to constrain the impact of environmental quenching towards lower masses, where we would expect much clearer signatures of cluster-specific quenching. Large surveys such as LSST, Euclid, the 4MOST CHANCES \citep{Haines2023}, and MOONRISE \citep{Maiolino2020} are ideal, as they will have both deep and wide data, allowing for detailed population studies that can constrain environmental quenching processes for a much more representative sample of clusters than this work.

\section{Conclusions}
\label{Conc}

In this work, we measured the LFs and SMFs of passive red-sequence galaxies in four galaxy clusters at $0.8 < z < 1.3$ down to masses below $10^{10}$ M$_{\odot}$, selected from deep VLT observations and complemented by the GCLASS and GOGREEN surveys. Our aim was to investigate the role of environmental quenching in shaping galaxy properties at these epochs, focusing on low-mass galaxies, a regime where discrepancies between models and observations are most pronounced. By comparing the cluster populations to corresponding field samples, we sought to clarify whether environmental quenching processes differ significantly at higher redshifts compared to the local Universe.

Our results reveal a clear difference in the faint and low-mass slopes of the LFs and SMFs of passive galaxies between cluster and field environments. We find a relative enhancement of faint, low-mass passive galaxies in clusters, as indicated by a shallower slope in the faint and low-mass end in the cluster LFs and SMFs. This result is seen in both the LFs and SMFs for each cluster, showing the robustness of these results. We also create a composite SMF for which we again find an enhancement in the low-mass passive galaxies in clusters, where we measure the Schechter parameter $\alpha = -0.54^{+\,0.03}_{-0.03}$ for the clusters and $\alpha = 0.12^{+\,0.02}_{-0.02}$ for the field. This finding contrasts with results from previous studies at similar redshifts that suggest that traditional, post-infall quenching may not be occurring as expected at high redshift due to the almost identical SMF shapes measured between cluster and field environments. The relative excess we measure in the low-mass end of the cluster SMF may be indicative of early stages of environmental quenching mechanisms already operating by $z\sim1$. This is consistent with theoretical models predicting that the impact of environmental quenching becomes more conspicuous for lower-mass galaxies in dense environments. We determine that to reproduce the slope at the low-mass end of the passive cluster SMF, $25\pm5\%$ of the star-forming field population would need to be quenched and combined with the already passive field population.

Our results largely support traditional quenching models, though deeper studies of larger samples of clusters are needed to get a better understanding of the role of environmental quenching in the distant Universe.

\section*{Acknowledgements}
HG acknowledges support through an STFC studentship. NAH acknowledges support from STFC through grant ST/X000982/1, and gratefully acknowledges support from the Leverhulme Trust. BF acknowledges support from JWST-GO-02913.001-A. BV acknowledges support from the INAF Mini Grant 2022, “Tracing filaments through cosmic time” (PI Vulcani). DCB is supported by an NSF Astronomy and Astrophysics Postdoctoral Fellowship under award AST-2303800. DCB is also supported by the UC Chancellor's Postdoctoral Fellowship. R.D. gratefully acknowledges support by the ANID BASAL project FB210003. GR acknowledges support from the NSF through AAG grant 2206473. YMB acknowledges support from UK Research and Innovation through a Future Leaders Fellowship (grant agreement MR/X035166/1) and financial support from the Swiss National Science Foundation (SNSF) under project 200021\_213076. G.W. gratefully acknowledges support from the National Science Foundation through grant AST-2205189. This research was supported by the International Space Science Institute (ISSI) in Bern, through ISSI International Team project “Understanding the evolution and transitioning of distant proto-clusters into clusters.” We are grateful for the support of ISSI and the use of their facilities. We are also grateful to the reviewer for their prompt and insightful feedback.

For the purpose of open access, the author has applied a creative commons attribution (CC BY) to any author accepted manuscript version arising.

\section*{Data Availability}

The GCLASS and GOGREEN data is publicly available at \url{https://www.canfar.net/storage/vault/list/GOGREEN/DR1}. The COSMOS2020 data can be accessed at \url{https://cosmos2020.calet.org/}. The PAU data is available at \url{https://pausurvey.org/}. The VLT data presented in section~\ref{VLTObs} is available upon request.



\bibliographystyle{mnras}
\bibliography{refs} 




\appendix



\bsp	
\label{lastpage}
\end{document}